\documentclass[pdflatex,sn-nature]{sn-jnl}% Math and Physical Sciences Reference Style
\usepackage{graphicx}%
\usepackage{multirow}%
\usepackage{amsmath,amssymb,amsfonts}%
\usepackage{amsthm}%
\usepackage{mathrsfs}%
\usepackage[title]{appendix}%
\usepackage{xcolor}%
\usepackage{textcomp}%
\usepackage{manyfoot}%
\usepackage{booktabs}%
\usepackage{algorithm}%
\usepackage{algorithmicx}%
\usepackage{algpseudocode}%
\usepackage{listings}%

\usepackage{cleveref}
\usepackage{caption}
\usepackage{subcaption}
\usepackage{upgreek}
\usepackage{aas_macros}
\usepackage{soul}
\usepackage{longtable}
\usepackage{nameref}

\begin{document}

\title[A Reverse Shock in GRB 221009A]{Precise Measurements of Self-absorbed Rising Reverse Shock Emission from Gamma-ray Burst 221009A}


\author*[1]{\fnm{Joe} S. \sur{Bright}}\email{joe.bright@physics.ox.ac.uk}
\equalcont{These authors contributed equally to this work.}

\author[1]{\fnm{Lauren} \sur{Rhodes}}
\equalcont{These authors contributed equally to this work.}

\author[2]{\fnm{Wael} \sur{Farah}}

\author[1,3]{\fnm{Rob} \sur{Fender}}

\author[4]{\fnm{Alexander} J. \sur{van der Horst}}

\author[5,6,7]{\fnm{James} K. \sur{Leung}}

\author[8]{\fnm{David} R. A. \sur{Williams}}

% alphabetical from here

\author[9]{\fnm{Gemma} E. \sur{Anderson}}

\author[10]{\fnm{Pikky} \sur{Atri}}

\author[11]{\fnm{David} R. \sur{DeBoer}\textsuperscript{11}}

\author[12,13]{\fnm{Stefano} \sur{Giarratana}}

\author[14]{\fnm{David} A. \sur{Green}}

\author[1,15,16]{\fnm{Ian} \sur{Heywood}}

\author[6]{\fnm{Emil} \sur{Lenc}}

\author[5,7]{\fnm{Tara} \sur{Murphy}}

\author[2]{\fnm{Alexander} W. \sur{Pollak}}

\author[2,17]{\fnm{Pranav} H. \sur{Premnath}}

\author[14]{\fnm{Paul} F. \sur{Scott}}

\author[2,18,19]{\fnm{Sofia} Z. \sur{Sheikh}}

\author[2,20,21,22]{\fnm{Andrew} \sur{Siemion}}

\author[14]{\fnm{David} J. \sur{Titterington}}


\affil*[1]{\orgdiv{Astrophysics, Department of Physics}, \orgname{The University of Oxford}, \orgaddress{\street{Keble Road}, \city{Oxford}, \postcode{OX1 3RH}, \country{UK}}}

\affil[2]{\orgname{SETI Institute}, \orgaddress{\street{339 Bernardo Ave}, \city{Suite 200 Mountain View}, \postcode{CA 94043}, \country{USA}}}

\affil[3]{\orgdiv{Department of Astronomy, University of Cape Town, Private Bag X3, Rondebosch 7701, South Africa}}

\affil[4]{\orgdiv{Department of Physics}, \orgname{The George Washington University}, \orgaddress{725 21st Street NW, Washington}, \postcode{DC 20052}, \country{USA}}

%\affil[5]{\orgdiv{Astronomy, Physics, and Statistics Institute of Sciences (APSIS), The George Washington University, Washington, DC 20052, USA}}

\affil[5]{\orgdiv{Sydney Institute for Astronomy}, \orgname{School of Physics}, \orgaddress{The University of Sydney}, \postcode{NSW 2006}, \country{Australia}}

\affil[6]{\orgdiv{CSIRO, Space \& Astronomy, PO Box 76, Epping, NSW 1710, Australia}}

\affil[7]{\orgdiv{ARC Centre of Excellence for Gravitational Wave Discovery (OzGrav), Hawthorn, Victoria, Australia}}

\affil[8]{\orgdiv{Jodrell Bank Centre for Astrophysics, School of Physics and Astronomy, The University of Manchester, Manchester, M13 9PL, UK}}

\affil[9]{\orgdiv{International Centre for Radio Astronomy Research}, \orgname{Curtin University}, \orgaddress{GPO Box U1987, Perth}, \postcode{WA 6845}, \country{Australia}}

\affil[10]{\orgdiv{ASTRON}, \orgname{Netherlands Institute for Radio Astronomy}, \orgaddress{Oude Hoogeveensedijk 4}, \postcode{7991 PD Dwingeloo}, \country{The Netherlands}}

\affil[11]{\orgdiv{Radio Astronomy Lab}, \orgname{University of California, Berkeley, CA}, \country{USA}}

\affil[12]{\orgdiv{Department of Physics and Astronomy}, \orgaddress{University of Bologna, via Gobetti 93/2}, \postcode{40129 Bologna}, \country{Italy}}

\affil[13]{\orgdiv{INAF Istituto di Radioastronomia}, \orgaddress{via Gobetti 101}, \postcode{40129 Bologna}, \country{Italy}}

\affil[14]{\orgdiv{Astrophysics Group}, \orgname{Cavendish Laboratory}, \orgaddress{19 J. J. Thomson Avenue, Cambridge}, \postcode{CB3 0HE}, \country{UK}}

\affil[15]{\orgdiv{Department of Physics and Electronics, Rhodes University, PO Box 94, Makhanda 6140, South Africa}}

\affil[16]{\orgdiv{South African Radio Astronomy Observatory (SARAO), 2 Fir Street, Observatory, Cape Town 7925, South Africa}}

\affil[17]{\orgdiv{Department of Physics \& Astronomy, The University of California, Irvine, Irvine, CA 92697, USA}}

\affil[18]{\orgdiv{Berkeley SETI Research Center, University of California, Berkeley, CA 94720, USA}}

\affil[19]{\orgdiv{Penn State Extraterrestrial Intelligence Center, 525 Davey Laboratory, The Pennsylvania State University, University Park, PA, 16802, USA}}

\affil[20]{\orgdiv{Department of Physics and Astronomy, University of Manchester, UK}}

\affil[21]{\orgdiv{University of Malta, Institute of Space Sciences and Astronomy, Msida, MSD2080, Malta}}

\affil[22]{\orgdiv{Breakthrough Listen,  University of California Berkeley, Berkeley CA 94720}}


\abstract{\textbf{The deaths of massive stars are sometimes accompanied by the launch of highly relativistic and collimated jets. If the jet is pointed towards Earth, we observe a ``prompt'' gamma-ray burst due to internal shocks or magnetic reconnection events within the jet, followed by a long-lived broadband synchrotron afterglow as the jet interacts with the circum-burst material. While there is solid observational evidence that emission from multiple shocks contributes to the afterglow signature, detailed studies of the reverse shock, which travels back into the explosion ejecta, are hampered by a lack of early-time observations, particularly in the radio band. We present rapid follow-up radio observations of the exceptionally bright gamma-ray burst GRB 221009A which reveal an optically thick rising component from the reverse shock in detail, both temporally and in frequency space. From this, we are able to constrain the size, Lorentz factor, and internal energy of the outflow while providing accurate predictions for the location of the peak frequency of the reverse shock in the first few hours after the burst. These observations challenge standard gamma-ray burst models describing reverse shock emission.}}


\maketitle

\section*{Introduction}\label{sec:intro}
Long Gamma-ray bursts (LGRBs), flashes of gamma-rays lasting a few to hundreds of seconds \cite{kouveliotou1993}, are well associated with jets launched during the violent deaths of massive stars (\cite{galama1998}, or see \cite{levan2016} for a recent review of GRB progenitors). Prompt gamma-ray emission is thought to be the result of internal jet processes, possibly colliding shells or magnetic reconnection events within a highly relativistic and collimated jet with a small inclination angle with respect to the Earth \cite{rees1994,kobayashi1997,granot2011}. The prompt emission is succeeded by a broadband afterglow of predominantly synchrotron radiation, associated with the jet ploughing into the circum-burst medium (CBM), which is affected by the mass loss history of the progenitor star. This interaction leads to the formation of at least two shocks, a forward shock (FS) that travels into the CBM and a reverse shock (RS) that propagates back into the GRB ejecta.

The afterglow emission from LGRBs is usually observed from radio through gamma-ray frequencies \cite{perley2014,vanderhorst2014,laskar2013}, and interpreted in the context of the ``fireball'' model \cite{piran1999}. The spectrum is described by a series of power-law segments separated by characteristic break frequencies normalised to some peak flux density \cite{granot2002}. The spectral breaks correspond to the frequency below which synchrotron self-absorption becomes dominant ($\nu_{sa}$), the emitting frequency of the lowest energy electrons in the shock accelerated population ($\nu_{m}$; ignoring any contribution from unaccelerated electrons \cite{ressler2017}), and the frequency above which electrons rapidly cool through emitting radiation ($\nu_{c}$). The evolution of these characteristic frequencies and the peak flux density are dependent on the density profile of the CBM ($\rho(r)\propto r^{-k}$) where $k=0$ characterises a homogeneous CBM and $k=2$ characterises a stellar-wind CBM (see \cite{granot2002, vanderhorst2014} for a comprehensive breakdown of these scaling relations). Due to the presence of (at least) two shock components, the broadband afterglow is formed of a superposition of (at least) two spectral components, evolving independently. Tracking the evolution of the two components is a powerful tool for understanding the jets powered by the death of massive stars, their interaction with the CBM, and the structure of the CBM itself.

The isotropic equivalent gamma-ray energy distribution of LGRBs spans almost seven orders of magnitude (between $\approx10^{48}\,\rm{erg}$ and $\approx10^{55}\,\rm{erg}$) with the more intrinsically luminous events detectable out to at least redshift $\approx9$ \cite{cucchiara2011, perley2014}. Long GRBs observed with the Neil Gehrels \textit{Swift} Observatory (hereafter \textit{Swift}) are most commonly found at a redshift $\approx2$ \cite{fynbo2009}, with over 80\% having redshifts $\gtrsim1$, and a narrower range of isotropic equivalent energies between $\approx10^{52}\,\rm{erg}$ and $\approx10^{53}\,\rm{erg}$ \cite{perley2014}. Such events are expected to be significantly less common in the local ($z\lesssim0.5$) Universe due to a reduction in star formation rate and the sensitivity of LGRB rates to metallicity \cite{robsertson2012,kruhler2015,pescalli2016,matthews2021}. In the local Universe, LGRB detections are dominated by low isotropic energy equivalent events, which could represent the tail end of the luminosity distribution of a single population not detectable at large distances, or a distinct population of low luminosity LGRBs \cite{soderberg2006,stanek2006,guetta2007}. Occasionally, however, a cosmological LGRB explodes in the local Universe and allows for precision testing of afterglow models \cite{vanderhorst2008,anderson2014,perley2014,vanderhorst2014,laskar2013,bright2019}. 

At UT 13:16:59 on 9\textsuperscript{th} Oct. 2022 (MJD 59861.5535, which we define as $T_{0}$) the Fermi Gamma-ray Burst Monitor \cite{meegan2009} detected an exceptionally bright burst (GRB 221009A; \cite{veres2022}). The \textit{Swift} Burst Alert, Ultra-Violet and Optical, and X-ray Telescopes started observing approximately one hour later, providing arcsecond localisation \cite{dichiara2022}. Initially identified as a Galactic transient and named Swift J1913.1$+$1946, GRB 221009A was then localised to a host galaxy at redshift $z=0.151$ \cite{deugartepostigo2022,castrotirado2022}. The isotropic equivalent energy estimate measured by Konus-Wind was $E_{\rm{iso}}\approx3\times10^{54}\,\rm{erg}$ \cite{frederiks2022} which is very energetic but not atypical for a LGRB (see figure 1 from \cite{perley2014}, and \cite{williams2023}). The combination of close proximity and high isotropic equivalent energy make GRB 221009A an extremely rare (e.g., \cite{williams2023, laskar2023, oconnor2023}) and bright event, prompting extensive follow-up at all wavelengths (e.g., \cite{2022ATel15664....1I, 2022GCN.32645....1B,2022GCN.32646....1D, 2022GCN.32652....1B,  2022GCN.32659....1P, 2022GCN.32669....1V, 2022GCN.32676....1D, 2022GCN.32678....1G}).

We report rapid radio follow-up observations of GRB 221009A with the Arcminute Microkelvin Imager Large Array (AMI--LA; \cite{zwart2008, hickish2018}) and Allen Telescope Array (ATA), beginning just $3.1\,\rm{hr}$ post burst as part of a larger radio monitoring campaign (including observations with \textit{e}-MERLIN, SMA, and ASKAP; see Methods Section `\nameref*{sub: AMI}' for details on all of our radio observations) which will be presented in full in later work. Due to the brightness of GRB 221009A and our fast response time, we captured rising optically thick emission from the reverse shock, and resolve variability on 15-minute timescales (\Cref{fig:total_light_curves}). Our ongoing radio campaign represents the best-sampled, multi-frequency, early-time look at a LGRB afterglow to date, and provides insights into the nature of the early-time reverse shock never previously possible. These observations constitute an important data set for the study of LGRB afterglows and the planning of future observing campaigns with the goal of understanding reverse shock emission.

\section*{Results}\label{sec:results}
\begin{figure}[H]
        \includegraphics[width=0.9\textwidth]{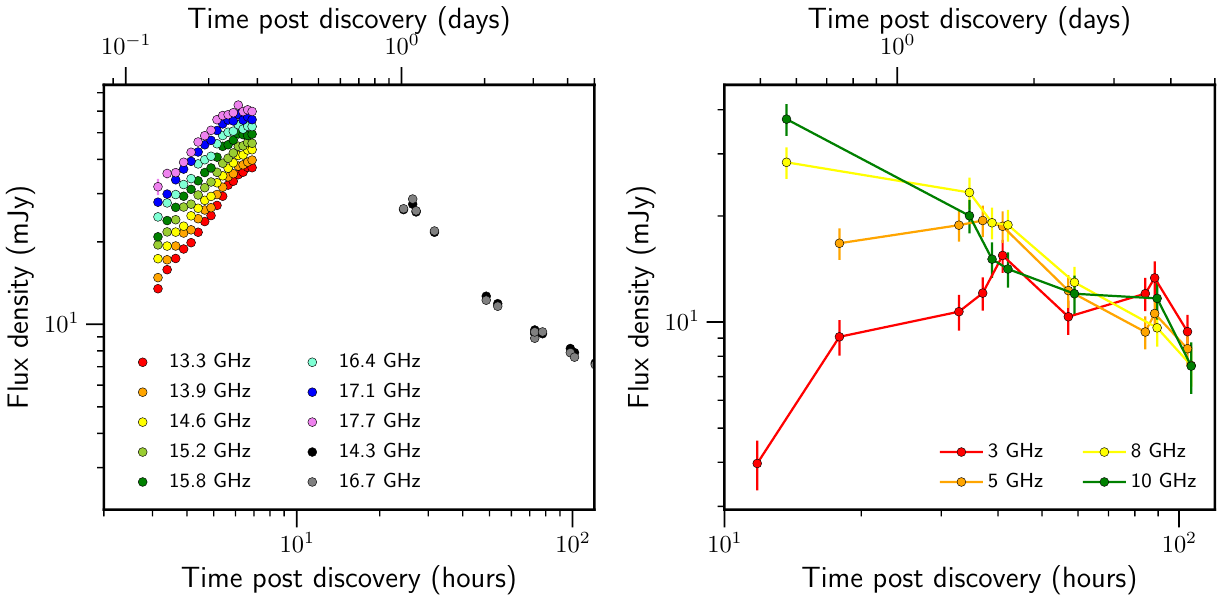}
        \caption{\textbf{Radio light curves of GRB 22109A between $3$ and $18\,\textrm{GHz}$.} \textbf{(left)} AMI--LA observations of GRB 221009A for the first 5 days post burst. Due to the high flux density in the first observation (between $T_{0}+3.1\,\rm{hr}$ and $T_{0}+7.1\,\rm{hr}$) we are able to split the data into $15\,\rm{min}$ time bins for each of the eight quick look spectral windows and derive flux density values directly from the complex visibilities. After the first day we derive fluxes from the image plane, using the top and bottom half of the AMI--LA observing band to monitor any spectral index evolution. See Methods for details on the data reduction process. \textbf{(right)} ATA observations of GRB 221009A for the first 5 days post burst showing an early-time peak most evident at $3$ and $5\,\rm{GHz}$ (and tentatively seen at $8\,\rm{GHz}$). All flux densities are derived from the image plane, see Methods for details on the data reduction process and imaging creation and processing. Error bars represent $1\sigma$ uncertainties.}
        \label{fig:total_light_curves}
\end{figure}

\begin{figure}[H]
    \centering
    \includegraphics[width=0.9\textwidth]{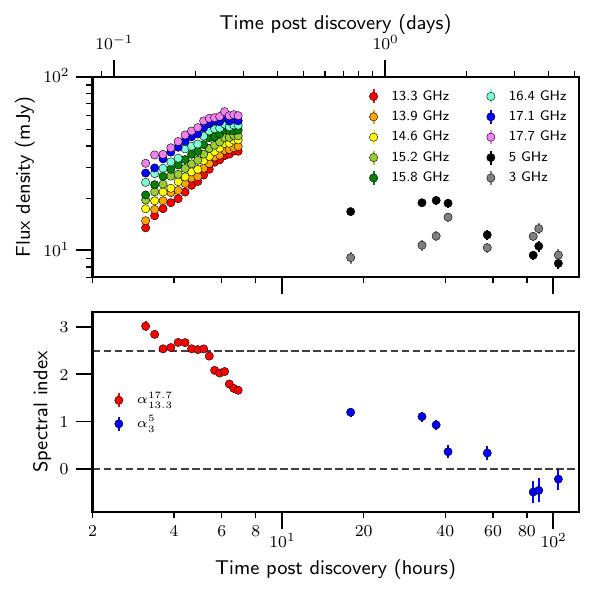}
    \caption{\textbf{Early-time radio light curves and spectral index evolution of GRB 221009A.} \textbf{Top:} Our first observation of GRB 221009A with the AMI--LA, separated into eight frequency channels and the flux density derived in $15\,\rm{min}$ time intervals as well the $3$ and $5\,\rm{GHz}$ light curves from the ATA. \textbf{Bottom:} The two-point spectral index $\alpha_{\nu_{1}}^{\nu_{2}}$ ($F_{\nu}\propto\nu^{\alpha_{\nu_{1}}^{\nu_{2}}}$, where $\nu_{1}$ and $\nu_{2}$ are the lower and upper frequencies used to calculate the two-point spectral index, respectively) measured between the highest and lowest of the eight AMI--LA quick-look frequency channels and the two ATA bands. Clear evolution can be seen throughout the observations, with the spectral index initially consistent with optically thick synchrotron ($\alpha\approx2.5$) and flattening with time. This is indicative of a break frequency (likely the self-absorption break) beginning to move into the AMI--LA observing band and then through the ATA observing bands. We mark the location of $\alpha=2.5$ and $\alpha=0$ with dashed horizontal lines to aid the reader. Error bars represent $1\sigma$ uncertainties.} 
    \label{fig:AMI_obs1}
\end{figure}

We initiated rapid follow-up radio observations of GRB 221009A with the AMI--LA (\cite{zwart2008,hickish2018}; \Cref{fig:total_light_curves}) and the ATA (\Cref{fig:total_light_curves}) beginning at $T_{0}+3.1\,\rm{hr}$ and $T_{0}+8.7\,\rm{hr}$, respectively (Extended Data Table \ref{tab:radio_data}). Observations with the AMI--LA were taken at a central frequency of 15.5\,GHz with a 5\,GHz bandwidth. Observations with the ATA were predominantly conducted at 3, 5, 8, and 10\,GHz. During the first observation with the AMI-LA we are able to separate the AMI-LA data into eight evenly spaced sub-bands, and into $15\,\rm{min}$ time bins, when measuring the flux density of GRB 221009A. We detect an exceptionally bright and rapidly rising radio counterpart to GRB 221009A (peaking at $\approx60\,\rm{mJy}$ at $17.7\,\rm{GHz}$ at $T_{0}+6.1\,\rm{hr}$). No archival radio source is evident at the position of GRB 221009A in wide-field radio surveys down to a 3$\sigma$ upper limit of 450\,$\mu$Jy (see Methods Section `\nameref*{sec:archival_rad_obs}'; Extended Data Figure \ref{fig:archival_radio}). Given the smoothly rising flux density in all eight of the AMI--LA bands, and a power law spectral index evolving smoothly with time (\Cref{fig:AMI_obs1}), we strongly disfavour scintillation as the cause of the early time variability (see Methods Section `\nameref*{sec:scint}').

In addition to the peak seen within the first 10 hours in all eight AMI--LA sub-bands, a clear peak is also seen in the 3 and $5\,\rm{GHz}$ light curves with the ATA, delayed by 34.3 and 17.7\,hr, respectively, compared to the peak observed at 17.7\,GHz (as well as a tentative peak at $8\,\rm{GHz}$). The observed behaviour is fully consistent with a spectral break moving to lower frequencies, through the AMI--LA band and through the four lower-frequency ATA bands, with time. The lack of a (clear) peak in the 8 and $10\,\rm{GHz}$ light curve indicates that the break frequency has already moved into/below the 8 and $10\,\rm{GHz}$ bands by $T_{0}+13.7\,\rm{hr}$ (the time of our first 8 and $10\,\rm{GHz}$ ATA observations). 

Due to our high cadence monitoring and wide frequency coverage, we are able to precisely track the evolution of the spectral peak. We fit a phenomenological model to our data assuming that the light curve can be described as a smoothly broken power-law and that the emission before $T_{0}+5\,\rm{d}$ is dominated by a single shock component (see Methods Section `\nameref*{sec:fitting}' and Extended Data Figure \ref{fig:fits} for the fitting procedure and results). We measure the rise and decay rate of the early-time radio emission to be $F_{\nu,\rm{rise}}\propto t^{1.34\pm0.02}$ and $F_{\nu,\rm{decay}}\propto t^{-0.82\pm0.04}$, respectively. Extended Data Table \ref{tab:fitting} contains the peak frequency and flux density at each of the frequencies where we measure the peak, and demonstrates the peak evolving from 17.7 to $3\,\rm{GHz}$ between 0.283 and $1.7\,\rm{d}$ post burst. Figure \ref{fig:rs_eval} shows the evolving frequency location of, and flux density at, the self-absorption peak. In the highest frequency band observed by the AMI--LA, we measure a fitted peak flux density of $57.2\pm0.6\,\rm{mJy}$, the brightest radio counterpart of any GRB detected to date.

In \Cref{fig:AMI_obs1} we show the spectral index evolution seen by the AMI--LA and the ATA, with the spectral index transitioning from self-absorbed ($F_{\nu}\propto\nu^{\sim2.5}$) to roughly flat ($F_{\nu}\propto\nu^{\sim0}$). The steep spectrum, sharp light curve rise, and peak timescale of less than one day implies that the peak of the light curves is a result of the synchrotron self-absorption break from the RS passing through the radio band. The flat spectrum that is measured post-peak is most likely a result of contamination from the forward shock as it enters the radio band, where superposition of the two synchrotron emission components can cause a flat spectral index.

\section*{Discussion \& Conclusions}\label{sec:dis_con}
\begin{figure}[H]
    \centering
    \includegraphics[width=0.9\textwidth]{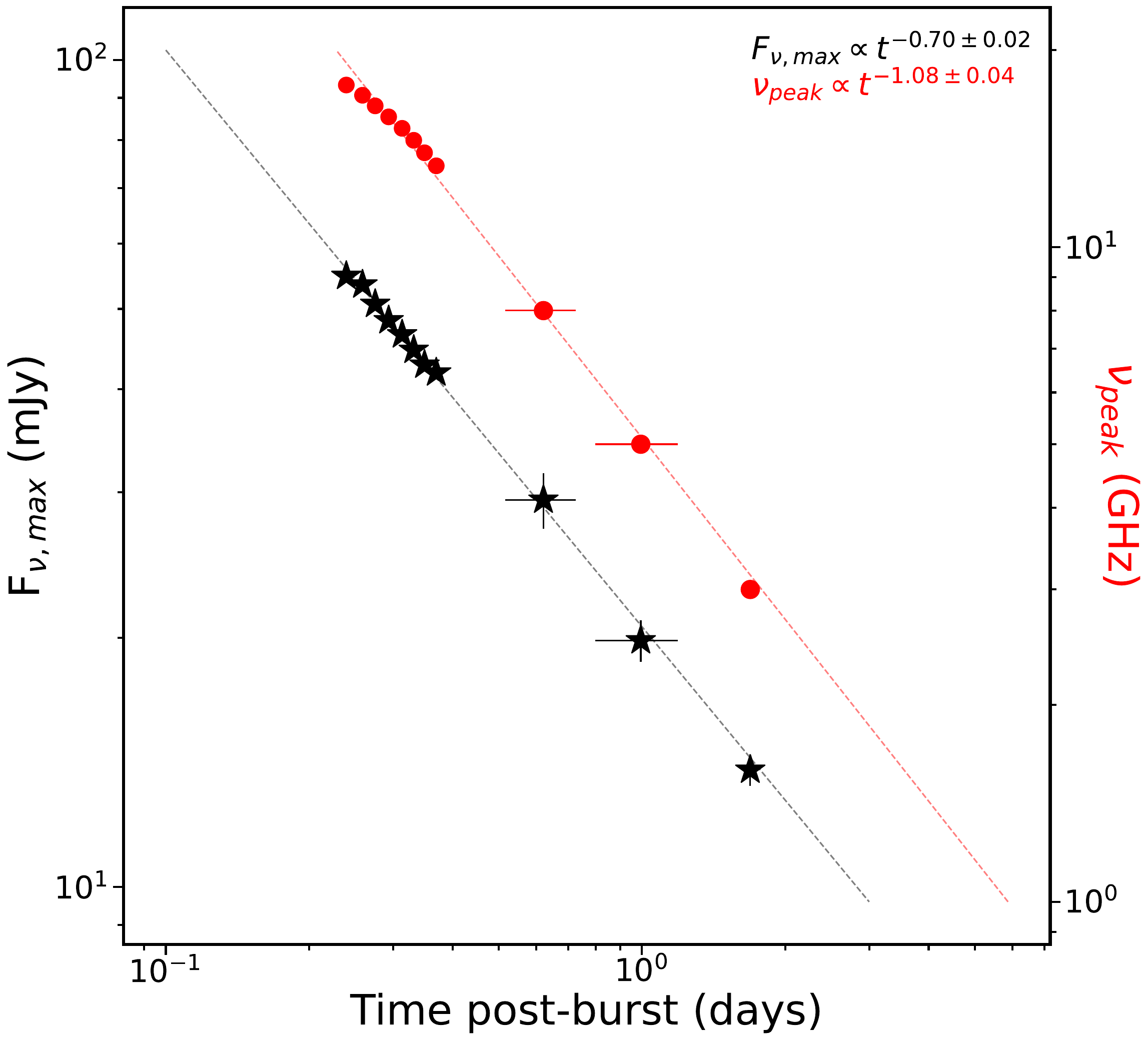}
    \caption{\textbf{Tracking the evolution of the peak frequency and time of the reverse shock in GRB 221009A.} The evolving frequency of, and flux density at, the early-time peak in our radio light curves of GRB 221009A. It can be seen to move to lower frequencies and flux densities with time. The peak flux density evolves as $F_{\nu,sa}\propto t^{-0.70\pm0.02}$ (shown as a dashed black line). The synchrotron self-absorption frequency evolves as $\nu_{sa}\propto t^{-1.08\pm0.04}$ (shown as a dashed red line). Error bars represent $1\sigma$ uncertainties.}
    \label{fig:rs_eval}
\end{figure}

\begin{figure}[H]
    \centering
    \includegraphics[width=0.9\textwidth]{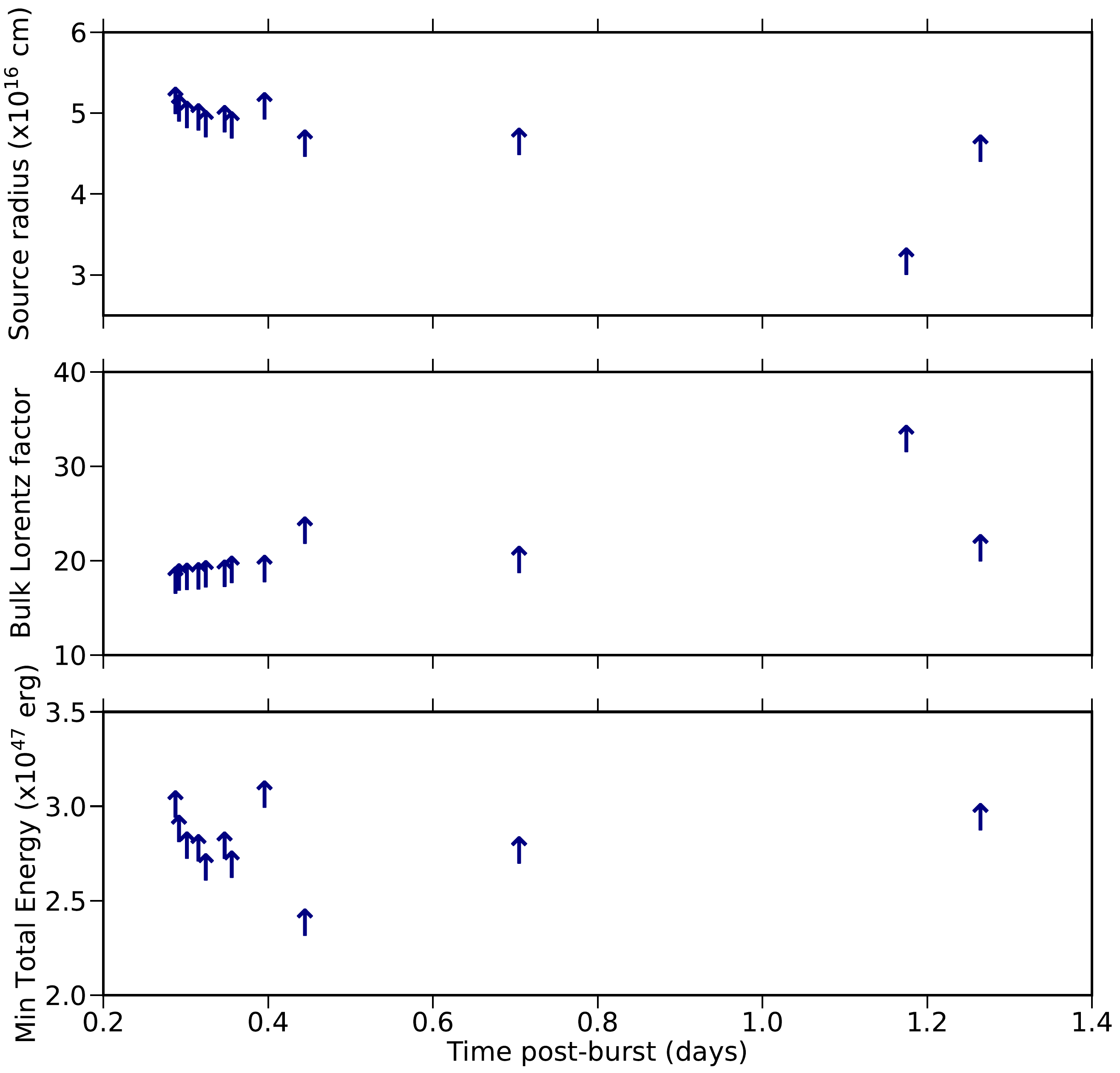}
    \caption{\textbf{An equipartition analysis of the self-absorbed emission from the reverse shock in GRB 221009A.} The results of an equipartition analysis using the peak flux densities, peak times, and observing frequencies from fitting power laws to the early time radio afterglows. Measuring the location of the peak allows us to constrain the emitting region size (top panel), bulk Lorentz factor (middle panel), and minimum total energy (bottom panel). These results are derived based on calculations described in Methods Section `\nameref*{subsubsec:eq_an}'.}% Error bars represent $1\sigma$ uncertainties.}
    \label{fig:bd13}
\end{figure}

The presence of the synchrotron self-absorption peak moving through the radio observing bands allows us to perform an equipartition analysis, and calculate constraints on the evolving size of the radio source, the bulk Lorentz factor, and minimum internal energy present within the RS of the GRB jet. Due to our rapid follow-up time, these constraints are among the earliest derived for any GRB. At $\approx T_{0}+6\,\rm{hr}$ we measure the size to be $\gtrsim5\times10^{16}$\,cm, the bulk Lorentz factor as $\gtrsim20$, and a minimum internal energy of the jet to be $\gtrsim3\times10^{47}\,\rm{erg}$. The full results of our equipartition analysis are shown in Figure \ref{fig:bd13}, and the method is described in detail in the Methods. It has been speculated that a jet break (the observational result of seeing the edge of a jet and the entirety of a jet becoming causally connected due to deceleration) is seen as an achromatic break in the X-ray and optical data at around $1\,\rm{d}$ post burst \cite{oconnor2023,levan2023}. Our Lorentz factor constraints at this time are consistent with the narrow opening angle ($\theta_{j}$) required for this jet break ($\Gamma\gtrsim20$ at $\sim T_{0}+1\,\rm{d}$ implies $\theta_{j}\lesssim3^{\circ}$) and inferred by \cite{williams2023}. We do not see evidence of a change in the evolution of the afterglow peak around the suggested jet break time (Figure \ref{fig:rs_eval} shows a smooth evolution of the spectral peak at all times), however as the radio emission is likely dominated by the RS at this stage a FS jet break cannot be ruled out from our data.

Our constraints on the source size ($\gtrsim10^{16}$\,cm) are comparable to those obtained through other methods including direct imaging, utilising very long baseline interferometry, and scintillation studies \cite{taylor2004,frail2005,alexander2019,anderson2022}. In order to reconcile the time of the peaks with the source size limits, we require an apparent expansion velocity of around 60$c$.
The requirement of unphysically high velocities is alleviated by the measurement of a significant Lorentz factor: our equipartition minimum for the Lorentz factor of the jet at 0.3 to 1\,days post burst sits above $\approx35$. This implies that the effects of superluminal expansion must be accounted for when considering the very high expansion velocities we infer. The constraints on the early-time Lorentz factor for GRB 221009A are broadly consistent with launch Lorentz factors derived for samples of LGRBs from broadband afterglow modelling or maximum brightness temperature arguments \cite{liang2010,liang2015,anderson2018}.

The limits we place on the internal energy via our equipartition and synchrotron self-absorption analysis are model-independent, whereas typically predictions are made in the context of broadband afterglow modelling. As we place constraints on both the Lorentz factor and minimum internal energy ($E_{i}$) we also place a lower limit on the total energy in the reverse shock component to be $E_{tot}=\Gamma E_{i}\gtrsim6\times10^{48}\,\rm{erg}$ at $T_{0}+6\,\rm{hr}$. Afterglow modelling in the context of the fireball model provides isotropic equivalent kinetic energies (i.e. not opening angle corrected) in the range of $10^{52}$ to $10^{54}\,\rm{erg}$, significantly higher than our equipartition estimate \cite{sari1998,chevalier2000,yost2003,aksulu2022}. Correcting the isotropic equivalent kinetic energy values for the jet geometry can reduce the results significantly. Additionally, while our analysis places a firm lower limit on the total energy, even modest departures from equipartition have significant implications for the total energy contained in the jet. Multi-wavelength modelling performed on GRB afterglows has shown that the energy distribution between the electrons and magnetic field is potentially significantly out of equipartition \cite{aksulu2022}. 

The first hours post-burst is a time frame that is rarely observed by radio facilities, and never before with such dense temporal and frequency coverage \cite{anderson2018}. Due to the rapid deceleration of the jet, the earlier we can obtain radio and sub-mm observations, ideally within the first few hours post-burst, the stricter the constraints that can be placed on the early-time properties of the LGRB jet. Our precise identification of the movement of the synchrotron self-absorption frequency and flux density means we can estimate the peak (sub-)mm flux density that would have been measured from GRB 221009A to be $\approx370\,\rm{mJy}$ at 230\,GHz and $T_{0}+0.4\,\rm{hr}$. Whilst a GRB as bright as GRB 221009A is rare, in order to discover early sub-mm counterparts to help determine the rarity and properties of self-absorbed reverse shocks, we require facilities such as the African Millimeter Telescope which plans to be able to slew to \textit{Swift}-detected bursts within minutes. 

We have been able to identify the location of (and flux density at) the synchrotron self-absorption frequency of the RS with high precision. We can use this information to make inferences about the structure of the RS and the density profile of the CBM. The evolution of the emission from the reverse shock depends strongly on the thickness of the ejecta material. The two extreme cases are for a \textit{thin-shell} ejecta, where the reverse shock quickly crosses the ejecta material and does not become relativistic (also known as a \textit{Newtonian} reverse shock), and for a \textit{thick-shell} ejecta, where the crossing time is significant and the shock velocity becomes relativistic (also known as a \textit{relativistic} reverse shock). For a \textit{Newtonian} reverse shock the observed evolution depends on the Lorentz factor distribution of the ejecta material ($\Gamma(r)\propto r^{-g}$; \cite{rees1998}) whereas for a \textit{relativistic} reverse shock the density profile of the CBM ($\rho \propto r^{-k}$; \cite{granot2002}) governs the evolution, and different evolution is predicted pre- and post-shock crossing.

The thickness of the RS can be constrained based on the rise and decay rate of the RS synchrotron emission, as well as from the evolution of the peak frequency (and the corresponding flux density) in a way that is outlined in \Cref{sec:fireball} of the Methods. We measure a power-law rise rate with an index of $1.34\pm0.02$, which corresponds to $k=1.5\pm0.1$ for a thick-shell RS and $g=1.7\pm0.1$ for a thin-shell RS. Both of these values are consistent with those expected from theoretical predictions of CBM density profiles ($k=[0,2]$) and jet structure ($g=[1/2,7/2]$), respectively. The decay rate of the radio flux density once the self-absorption break has moved below our observing band is not consistent with a wind-like environment, however FS contamination could be altering the evolution.

In addition to the post-break rise and decay rates, we also measure the movement of the peak of the reverse shock component in both flux density and frequency space to be $F_{\nu, max, sa}\propto t^{-0.70\pm0.02}$ and $\nu_{sa}\propto t^{-1.08\pm0.04}$ (\Cref{fig:rs_eval}), which can be compared with theoretical predictions for a reverse shock in the context of the thin and thick shell cases (see Methods Section `\nameref*{sec:fireball}'). In the thick shell regime, for a reasonable range of values $p=[1,4]$ and $k=[0,2]$, $F_{\nu, max, sa}$ is expected to decay at a rate between $t^{-1}$ and $t^{-1.9}$. In the thin shell regime, for $p=[1,4]$ and $g=[1/2,7/2]$, the peak is expected to decay with a rate between $t^{-0.9}$ and $t^{-2.2}$. Both scenarios are inconsistent with our observations, as we measure the peak to decay significantly slower than both of these cases. For thick and thin shell reverse shock models we expect the self-absorption break to decay at a rate between $t^{-0.9}$ and $t^{-1.3}$, and $t^{-0.8}$ and $t^{-1.6}$, respectively. The evolution of $\nu_{sa}$ is consistent with the slowest model decay rates for both thick and thin shell reverse shocks. Overall, we find that the light curve rise rate and the evolution of $\nu_{sa}$ are consistent with both thin and thick shell scenarios, however the post peak decay rate and evolution of the peak flux density are inconsistent with either scenario. It is possible that even at relatively early times ($T_{0}\lesssim1\,\rm{d}$) the FS is contributing significantly at radio frequencies.

\newpage

\backmatter

\setcounter{page}{1}

\bmhead{Methods}

\section{Archival Radio Observations}\label{sec:archival_rad_obs}
The field of GRB 221009A has been observed as part of the National Radio Astronomy Observatory Very Large Array Sky Survey (NVSS) and Very Large Array Sky Survey (VLASS) at $1.4\,\rm{GHz}$, and $3\,\rm{GHz}$, respectively. Using the Canadian Institute for Radio Astronomy Data Analysis image cutout web server we accessed pre-burst images of the field of GRB221009A. Extended Data \Cref{fig:archival_radio} shows both the VLASS and NVSS images of the field, where we do not find any significant archival emission at the location of GRB 221009A. From the VLASS image we place a three-sigma upper limit of $\sim450\,\upmu\rm{Jy/beam}$ at the location of GRB 221009A. Deep late-time observations might reveal faint host galaxy emission once the afterglow has faded. Two nearby field sources are evident in Extended Data \Cref{fig:archival_radio} and are also identified in our AMI--LA and ATA observations (see Supplementary Figures \ref{fig:ami_field} and \ref{fig:ata_field}, and the following section).

\section{Observations}
\subsection*{Arcminute Microkelvin Imager Large Array}\label{sub: AMI}

We began observations of the field of GRB 221009A on 09/10/2022 at UT 16:25:25.5 (MJD 59861.6843) with the Arcminute Microkelvin Imager Large Array (AMI--LA; \cite{zwart2008, hickish2018}). All observations are conducted at a central frequency of $15.5\,\rm{GHz}$ across a $5\,\rm{GHz}$ bandwidth consisting of 4096 frequency channels. In order to reduce the volume of recorded data and subsequent processing overheads we work with a `quick-look' data format where data are averaged into 8 broad and equivalent width frequency channels. Data are phase reference calibrated using the custom reduction software \textsc{reduce\_dc} with 3C 286 used to calibrate the bandpass and absolute flux scale of the array, while J1925$+$2106 is used to calibrate the time-dependent phases. We perform additional flagging and imaging in the Common Astronomy Software Applications (\textsc{casa} v4.7.0; \cite{mcmullin2007, thecasateam2022}) package using the \textit{rflag}, \textit{tfcrop}, and \textit{clean} tasks. We measure the flux density in the image plane using the \textit{imfit} task using two Taylor terms ($\textsc{nterms}=2$), splitting the observing bandwidth into equal halves after our first observation. The typical angular resolution of the AMI--LA is between $30''$ and $60''$ depending on the source declination and exact timing of the observation.

At early times when GRB 221009A was both bright and rapidly varying we extract the flux densities directly from the complex visibilities rather than from the image-plane. To do this we averaged the real part of the complex visibilities within $15\,\rm{min}$ time intervals, and set the error as the standard deviation of these amplitudes. While extracting flux densities in this way is strictly only correct for a point source at the phase center, with no other emission in the field, the flux density from GRB 221009A constituted $\gtrsim95\%$ of the total flux density in the image during this first observation, and flux densities measured in the image plane on longer (per-hour) timescales show good agreement with those derived from the visibilities directly. We give the short-timescale radio flux densities in Supplementary Data Table \ref{tab:ami_short}.

\subsection*{Allen Telescope Array}\label{sub:ATA}
Hosted at the Hat Creek Radio Observatory in Northern California, the Allen Telescope Array (ATA) is a 42-element radio interferometer that has been undergoing refurbishment since late 2019 aimed at improving the design and sensitivity of the telescope feeds, and at upgrading the digital signal processing (DSP) system (Pollak et al., \textit{in prep}.). Each ATA dish is $6.1\,\rm{m}$ in diameter and is fully-steerable with an offset-Gregorian design. The newly-refurbished, cryogenically-cooled dual-polarization log-periodic feeds are sensitive to a broad range of radio frequencies, 1-10 GHz \cite{welch2017}. Up to 4 independent frequency tunings, each $\sim700$\,MHz wide, can be selected anywhere in the RF band, theoretically allowing observers to tap into $\sim2.8$\,GHz of instantaneous bandwidth.

The currently-deployed DSP setup can digitise and process 2 tunings from 20 dual-polarization antenna feeds. The data are passed to a real-time xGPU-based \citep{xGPU} software correlator to produce a visibility dataset that can be subsequently imaged. Details about the digital signal processing chain will be presented in a subsequent paper (Farah et al., \textit{in prep}).

We reduced observations with the ATA using a custom pipeline implemented using \textsc{aoflagger} \cite{aoflagger} and \textsc{casa}. We flag the raw correlated data (which had a correlator dump time of $10\,\rm{s}$ for all frequencies) using \textsc{aoflagger} and default parameters before averaging the data in frequency by a factor of 8 (moving from a channel width of 0.5\,MHz to 4\,MHz) to reduce the processing and imaging time. Such averaging causes minimal bandwidth smearing for the field of view of the ATA at all frequencies and will not cause issues when imaging. We observe 3C 286, 3C147, or 3C48 for $10\,\rm{min}$ at the start of each observing session and used these observations to calibrate the bandpass response and absolute flux density scale of the array. We interleave a $10\,\rm{m}$ observation of J1925$+$2106 for every $30\,\rm{m}$ on the science target (at all frequencies) in order to calibrate the time-dependent phases. Total time on source varied throughout our observing campaign as we adjusted our observing strategy based on the brightness of GRB 221009A. We perform imaging using the \textsc{casa} task \textit{tclean} and a Briggs robust parameter \cite{briggsphd} of either 0 or 0.5, which we found to be a good compromise between sensitivity and restoring beam shape allowing for complete deconvolution. The inner region of an example ATA image, at a central frequency of 5\,GHz, is shown in Supplementary Figure \ref{fig:ata_field}. In order to extract the flux density associated with GRB 221009A we use the \textsc{casa} task \textit{imfit} where we fix the shape and orientation of the component to match the dimensions of the restoring beam for each observation. We constrained the fit to a small region around the source and fixed the location of the component to the brightest pixel consistent with the position of GRB 221009A when the contribution from nearby field sources became significant. The flux densities measured with the ATA are given in Extended Data Table \ref{tab:radio_data}. A full imaging pipeline is being developed for the ATA, and will be outlined in Farah et al. (\textit{in prep.}).

\subsection*{\textit{enhanced}-Multi-Element Radio Linked Interferometer Network}

We obtained observations with the \textit{enhanced}-Multi-Element Radio Linked Interferometer Network (\textit{e}-MERLIN) through an open time call proposal (CY14001, PI: Rhodes) and Rapid Response Time requesting (RR14001, PI: Rhodes). The data included in this article were taken at 1.51\,GHz, with a bandwidth of 520\,MHz. 

Observations began on 11-Oct-2022 at UT 17:55:32.5 and consisted of 8-minute scans on the science target, interleaved with 3-minute scans of the complex gain calibrator J1905+1943. The observation was bookended with a scan of each of the bandpass and flux calibrators, J1407+2827 and 3C 286, respectively. The data were reduced using the \textit{e}-MERLIN \textsc{casa}-based pipeline (Version 5.8; \cite{moldon2021}). The pipeline flags for radio frequency interference, performs bandpass calibration, and calculates amplitude and phase gain corrections which it then applies to the target field. We perform interactive cleaning and deconvolution using the task \textit{tclean} within \textsc{casa}. The final flux density of the source associated with GRB 221009A is given in Extended Data Table \ref{tab:radio_data}.

\subsection*{Australian Square Kilometre Array Pathfinder}

We obtained target-of-opportunity observations of the GRB 221009A field with the Australian Square Kilometre Array Pathfinder (ASKAP; \cite{johnston2007}), a wide-field radio telescope with a nominal field-of-view of ${\sim}30$ square degrees. Our observations were centred on 888 MHz, with a bandwidth of 288 MHz, taken using the \texttt{square\_6x6} beam footprint (see figure 20 of \cite{hotan2021}). The observations were pointed at $\rm{RA}=19{:}20{:}00.00$, $\rm{Dec}=+20{:}18{:}05.0$, placing the sky direction of GRB 221009A at the centre of beam 14. The observation (SB44780) included in this work began on 2022 October 12 at 07:02 UTC, with a total duration of 6\,h.
 
Observations of PKS B1934$-$638 were used to calibrate the antenna gains, bandpass and the absolute flux density scale.
Flagging of radio frequency interference, calibration of raw visibilities, full-polarisation imaging and source finding on total intensity images were all performed through the standard \textsc{ASKAPsoft} pipeline \citep{guzman2019}.
The resulting image has an rms of 47\,$\mu$Jy$/$beam and a $16.4''$ by $12.5''$ resolution.
The flux density scale of field sources evaluated against the Rapid ASKAP Continuum Survey (RACS) catalogue \citep{hale2021} is given by $S_{\textrm{ASKAP}}/S_{\textrm{RACS}} = 1.05 \pm 0.12$ and this systematic flux offset is accounted for in the measurement we report in Extended Data Table \ref{tab:radio_data}.

\section{Fitting}\label{sec:fitting}

The multi-frequency light curves are best described using a broken power law:

\begin{equation}
    F(t)=A\bigg[\frac{1}{2}\Big(\frac{t}{t_{b}}\Big)^{-sa_{1}}+\frac{1}{2}\Big(\frac{t}{t_{b}}\Big)^{-sa_{2}}\bigg]^{-1/s}
\end{equation}

which describes a smooth transition between two power-laws ($a_1$ and $a_2$) at $t\approx t_{b}$. Given that we are describing light curves that rise and then decay we can take, without loss of generality, $a_2<0<a_1$ such that in the limit $t\ll t_b$; $F(t)\propto t^{a_1}$, and in the limit $t\gg t_b$; $F(t)\propto t^{a_2}$. The parameter $s$ describes how smooth the transition is between the two extremes. We fix $s=2$ when fitting our light curves. We fix the power-law indices to be constant across the different bands whereas the peak flux densities and times are allowed to vary for each frequency. To better constrain the early time peak, we fit the light curve rise simultaneously across all eight AMI--LA sub-bands with a power-law (we do not expect any significant contribution from the FS component at the time of the AMI--LA rise \cite{zhang2004}) and use that in the broadband modelling. The joint fit to the AMI--LA data has a temporal index of $t^{1.34\pm0.02}$. From the broadband light curve fits, we find a decay rate of $t^{-0.83\pm0.02}$. The peak times and flux densities are shown in Extended Data Table \ref{tab:fitting}. The complete results of the model fitting will be presented in Rhodes et al. (\textit{in prep.}). 
%We fit each of the light curves simultaneously with the function

%The flux densities and times calculated are then used in the equipartition analysis described in Section \ref{subsubsec:eq_an}.
\section{Scintillation}\label{sec:scint}

Scintillation is a process by which apparently random fluctuations in radio light curves and spectra occur as a result of the diffraction or refraction of radio waves as they pass through the interstellar medium of the Milky Way. Scintillation can be broadly split into weak and strong regimes, divided by some transition frequency. The transition frequency is dependent on the levels of turbulence along the line of sight, characterised by the \textit{scattering measure}, where closer to the galactic plane/centre the scattering measure is higher and therefore the transition frequency is also higher \citep{rickett1990}. This means that the effects of strong scintillation can be observed up to higher frequency compared to sources far from the galactic plane (often the case for GRBs, however not for GRB 221009A which is observed through the plane of the Milky Way).

According to the NE2001 model for the Galactic distribution of free electrons \citep{cordes2002} the AMI--LA observing band ($15.5\,\rm{GHz}$) is firmly in the strong scintillation regime, in which flux density modulations of order 100\% could be observed. The model predicts a scintillation timescale of $\sim3\,\rm{min}$. Given our shortest integration time is 15\,min, we would expect to average out the effects of scintillation on the timescale of our observations. The variability we observe in the early time AMI--LA data consists of a monotonic increase in flux density (Figure \ref{fig:total_light_curves}). There is no random element to the variability that is characteristic of scintillation, and the spectrum evolves smoothly through values consistent with synchrotron radiation. Combined with the smoothly evolving spectral index measurements, we are confident in ruling out scintillation as the cause of the flux density changes.

\section{Inferences from X-ray data}\label{sec:x-ray}
X-ray observations of GRBs are uncontaminated by the reverse shock from very early times, and so we can attempt to use the spectral and temporal decay rates to constrain the physical properties of the FS, which can then be compared with those derived from the reverse shock. \textit{Swift}-XRT, MAXI, and NICER $0.3$-$10\,\rm{keV}$ data compiled by \cite{williams2023} demonstrate that the X-ray light curve is best described by a broken power law with the break occurring at $T_{0}+0.086\,\rm{d}$. Pre- and post-break the light curve decays as $F_{X}(t)\propto t^{-1.498\pm0.004}$ and $F_{X}(t)\propto t^{-1.672\pm0.008}$, respectively. The spectral index is $\sim-0.7$ ($F_{X}(\nu)\propto \nu^{-0.7}$). A corresponding (in time) break is not statistically preferred over a single broken power-law when describing the \textit{Swift-UVOT} data, which indicates that the X-ray break is unlikely to be the result of a jet break.

We expect both the self-absorption and minimum energy breaks to be well below the XRT observing band, regardless of their ordering, and so the only consideration is the location of the cooling break. For approximately the first day post-burst, spectral fitting shows that $\nu_{c}$ is consistent with lying in the \textit{Swift} XRT band \cite{williams2023}. During this period the spectral break is seen to move to lower energies (from $\sim6.8$\,keV to $\sim4.5$\,keV; see table 2 of \cite{williams2023}) indicating that after the first day post-burst, the X-ray band is above the cooling break, resulting in an expected spectral index of $-p/2$. In the aforementioned scenario (where $\nu_c$ lies below the XRT band) we would therefore obtain $p\approx1.4$ which is low but not unusually so for long GRBs \cite{curran2010}. This value of $p$, however, would dictate the X-rays decline with an index significantly shallower than observed (although the presence of a jet break would predict a decline rate of $-p$, which is more consistent with the observed rate). A similar conclusion was drawn by \cite{levan2023,laskar2023} based on a comparison with optical data. Conversely, if the X-ray band is below the cooling break, we obtain $p\approx 2.4$ (from the spectral index) which is in good agreement with population studies of X-ray emission from GRBs \cite{curran2010}. However, again, this would require a shallower decay index than observed in the ISM case but would be consistent with the measured decay in a wind scenario. It is worth noting that reconciling any of these scenarios with lower frequency (optical, NIR, and radio) data has proven to be challenging, leading authors to suggest extensions to the standard model, including structured jets \cite{levan2023,oconnor2023}. In summary, it appears that a standard forward shock model is not sufficiently constraining to confidently infer values for either $p$ or $k$ from X-ray observations alone.

\section{In the context of the fireball model}\label{sec:fireball}

Our temporal and spectral coverage in the radio band over the first three days post-burst allows us to test GRB reverse shock models in a way that has not been possible before. The spectral index measured across the 8 AMI-LA quick-look channels (see Figure \ref{fig:AMI_obs1}) shows a clear evolution from optically thick ($F_{\nu} \propto \nu^{2.5}$) to shallower values with time. The final spectral index measurement (from the last 15 minute interval) on the first night of AMI observations is $1.66\pm0.01$ (i.e. $F_{\nu} \propto \nu^{1.66}$). The optically thick spectral index (shown in the lower panel of Figure \ref{fig:AMI_obs1}) indicates that for the first 8\,hr post-burst the AMI-LA observing band is in the regime where $\nu_{m} < \nu_{obs} < \nu_{sa}$ (where $\nu_{m}$ is the minimum energy frequency and $\nu_{sa}$ is the self-absorption frequency). The spectral index post-peak is much flatter than expected from optically thin synchrotron (i.e. in the regime $\nu_{m} < \nu_{sa} < \nu_{obs}$). We expect $F_{\nu} \propto \nu^{\sim-0.7}$ but measure $F_{\nu} \propto \nu^{\sim-0.2}$ at 1\,day post-burst across the AMI--LA band, lasting until at least $5\,\rm{d}$ post burst as measured between 3 and $5\,\rm{GHz}$ with the ATA.

In the context of the fireball model \cite{piran1999}, the presence of (at least) two shock components, each with three time-dependent characteristic frequencies and a peak flux density, leads to a complex evolving broadband SED that can be hard to interpret without intense temporal and spectral coverage that spans weeks post burst and covers many orders of magnitude in frequency. This is further complicated by a range of possible values of the power-law index of the shock-accelerated electrons, $p$ ($N(E)\propto E^{-p}$), which produce distinct evolution in the cases of $1<p<2$ and $p>2$. Finally, the density profile of the CBM alters the evolution of the forward and reverse shock (which itself can either be thick or thin-shelled and evolves differently before and after the crossing timescale). As such, care must be taken to consider all possibly feasible scenarios when interpreting observational data. A comprehensive list of the evolution of the characteristic frequencies, the flux density at these frequencies, and the temporal and spectral evolution of the regions between them is given in \cite{gao2013,vanderhorst2014} with full consideration of the previously mentioned complications. From the spectral index measurements, we expect that the early time peak is caused by the transition from $\nu_{m} < \nu_{obs} < \nu_{sa}$ to $\nu_{m} < \nu_{sa} < \nu_{obs}$ and so we can check if any of the scenarios presented in \cite{gao2013,vanderhorst2014} are consistent with the observed rise rate, decay rate, and evolution of the break frequency and flux density (these presentations differ slightly in the regions of parameter space they cover for $p$, $k$, and $g$). We give a summary of the relevant scaling relations in Supplementary Data Table \ref{tab:relevant_scalings}.

The rise component of the radio light curve follows $t^{1.34\pm0.02}$. This behaviour can be explained in the thick shell regime for $k = 1.5\pm0.1$ (where $\rho(r) \propto r^{-k}$) and in the thin shell model for $g = 1.7\pm0.1$ (where $\Gamma(r) \propto r^{-g}$) which is valid for all $p>1$ in both cases (\cite{gao2013,vanderhorst2014}, and see Supplementary Data Table \ref{tab:relevant_scalings}). We note that the thin shell regime is relatively insensitive to the profile of the surrounding environment and is instead predominantly sensitive to the deceleration of the jet. With the rise alone we cannot distinguish between the thick and thin shell regimes. Post-peak we measure a light curve decay of $t^{-0.83\pm0.02}$, which now also depends on the accelerated electron energy index ($p$) in addition to $k$ and $g$. As a result, we can predict a range of decay values for $p>1$. In the thick shell regime, using $k=1.5$ as derived from the rise under the assumption of a thick shell gives a too steep decay for all $p>1$. In the thin shell model, using $g=1.7$ as derived from the rise under the assumption of a thin shell, we similarly find that no valid value of $p$ can explain the decay rate.

To further distinguish between different shell-thickness regimes, we use the broken power law fits from Section \ref{sec:fitting} to track the peak of the reverse shock spectrum from 18 to 3\,GHz. The peak flux density evolves as $F_{\nu,sa}\propto t^{-0.70\pm0.02}$ and the synchrotron self-absorption frequency evolves as $\nu_{sa}\propto t^{-1.08\pm0.04}$ (see Figure \ref{fig:rs_eval}). When deriving the evolution rate of the peak flux density under the assumption that it is caused by synchrotron self-absorption we make the assumption that $F_{max,\nu_{sa}}(t)=F_{max,\nu_{m}}(t)(\nu_{sa}(t)/\nu_{m}(t))^{(1-p)/2}.$ (as done in e.g., \cite{granot2002}) In the thick shell regime, the peak flux density of the reverse shock spectrum, if the peak is produced by $\nu_{sa}$, is expected to evolve as $t^{(-2k(12p+13)+126p+109)/(12(k-4)(p+4))}$ \citep{vanderhorst2014, gao2013}. For reasonable values of $p$ and $k$ (between $1-4$ and $0-2$, respectively), $F_{\nu, \rm{max}}$ is expected to decay at a rate between $t^{-1}$ to $t^{-1.9}$. In the thin shell regime $F_{\nu, \rm{max}}$ evolves as $t^{(-5g(5p+6)+20(2p+1))/(7(2g+1)(p+4))}$, which for reasonable values of $p$ and $g$ (between $1-4$ and $1/2-7/2$, respectively) corresponds to a decay rate between $t^{-0.9}$ and $t^{-2.2}$, inconsistent with out data.

Finally, we compare the evolution of $\nu_{sa}$ with theoretical predictions for reverse shock evolution. We find that there is a significant overlap between certain areas of parameter space when comparing the possible values of $\nu_{sa}(t)$ and our measured evolution. In the thick shell regime $\nu_{sa} \propto t^{-(p(73-14k)+2(67-14k))/(12(4-k)(p+4))}$ which corresponds to values between $t^{-0.9}$ and $t^{-1.3}$. For a thin-shell jet $\nu_{sa} \propto t^{-(3p(5g+8)+8(4g+5))/(7(2g+1)(p+4))}$ which spans a larger range of values between  $t^{-0.8}$ and $t^{-1.6}$. Given our measured value of $\nu_{sa}\propto t^{-1.08\pm0.04}$, the spectral evolution we obtained is possible within both thick and thin shell jets scenarios.

In summary, when considering the evolution of specific parts of the radio emission from the reverse shock (the rise rate, and evolution of the frequency of the self-absorption break) good agreements can be found with theoretical predictions. However, the decay rate and the evolution of the peak flux density are not recreated in either the thin- or thick-shell reverse shock models. We note that the post break temporal decay rate, and evolution of the peak flux density, might be explained by considering early time FS contamination.

\section{Equipartition analysis}\label{subsubsec:eq_an}

Following \cite{barniolduran2013}, we can place constraints on the physical parameters associated with the emitting region responsible for the early-time self-absorbed radio peak seen with the AMI--LA and ATA. This is possible under the assumption of equipartition and knowledge of the location of the synchrotron self-absorption frequency. The method results in robust lower limits on the radius, bulk Lorentz factor, and internal energy as

\begin{subequations}
    \begin{equation}    R_{eq}\approx(7.5\times10^{17}\,\textrm{cm})\bigg[F_{\rm{p,mJy}}^{\frac{2}{3}}d_{L,28}^{\frac{4}{3}}\nu_{p,10}^{-\frac{17}{12}}(1+z)^{-\frac{5}{3}}t_{d}^{-\frac{5}{12}}\bigg]f_{\textrm{A}}^{-\frac{7}{12}}f_{\textrm{V}}^{-\frac{1}{12}}
    \end{equation}

    \begin{equation}
    \Gamma\approx12\bigg[F_{\textrm{p,mJy}}^{\frac{1}{3}}d_{\textrm{L},28}^{\frac{2}{3}}\nu_{p,10}^{-\frac{17}{24}}(1+\textrm{z})^{-\frac{1}{3}}\textrm{t}_{d}^{-\frac{17}{24}}\bigg]f^{-\frac{7}{24}}_{\textrm{A}}f_{\textrm{V}}^{-\frac{1}{24}}
    \end{equation}

    \begin{equation}
    E_{eq}\approx(5.7\times10^{47}\,\textrm{erg})\bigg[F_{\textrm{p,mJy}}^{\frac{2}{3}}d_{L,28}^{\frac{4}{3}}\nu_{p,10}^{\frac{1}{12}}(1+z)^{-\frac{5}{3}}t_{d}^{\frac{13}{12}}\bigg]f^{-\frac{1}{12}}_{\textrm{A}}f_{\textrm{V}}^{\frac{5}{12}}.
    \end{equation}
\end{subequations}

The quantities within square brackets are observables, and we work under the assumption that the peak of the RS SED is due to the self-absorption break (we assume $\eta=1$ from \cite{barniolduran2013}; see main text for justification). These equations are valid in the case that $p>2$. Additionally, $F_{\rm{p,mJy}}$ is the flux density of the peak of the radio SED in units of mJy, $d_{L,28}$ is the luminosity distance in units of $10^{28}\,\rm{cm}$, $\nu_{p,10}$ is the frequency at which the SED peaks in units of $10\,\rm{GHz}$, and $t_{d}$ is the time at which the peak occurs measured in days. The geometry of the emitting region is encoded in $f_{A}$ and $f_{V}$ which are the fraction of the observed area and volume filled by the source, respectively. The equipartition radius, Lorentz factor, and energy are all only weakly dependent on $f_{\textrm{A}}$ and $f_{\textrm{V}}$ and as such we take $f_{\textrm{A}}=f_{\textrm{V}}=1$ in our analysis. At early times, the opening angle is greater than 1/$\Gamma$ and so we underestimate the total minimum energy by $4\Gamma^2 (1-\cos\theta_{j})$ where $\theta_{j}$ is the jet opening angle. Until  a jet break is observed it is not possible to calculate the opening angle and so we leave the minimum energy lower limits as is. 

The equipartition method presented in \cite{barniolduran2013} calculates the equipartition radius $R_{eq}$, which is actually the distance between the radio source and the launch site (see figure 1 from \cite{barniolduran2013}). In order to calculate the size of the GRB jet on the sky we have to transform $R_{eq}$ into the source radius using $R = R_{eq}/\Gamma$, as we are only able to observe radiation from within an opening angle of $1/\Gamma$. The radius estimates in Figure \ref{fig:bd13} include this correction. 

Finally, we consider the possibility that the GRB jet is not directly pointed along our line of sight, as is assumed in \cite{barniolduran2013}. In Supplementary Figure \ref{fig:lf_contour}, we re-parameterise the bulk Lorentz factor $\Gamma$ in terms of the Doppler factor $\delta$ and the angle to the line of sight to demonstrate the range of parameter space that can be explored if we drop the assumption that the angle to the line of sight is zero degrees.

\section{Implications for Future Observing Campaigns}
Due to our ability to constrain the peak of the reverse shock emission as a function of frequency and time we can make accurate predictions for the flux density at mm and sub-mm frequencies. This is particularly relevant when motivating rapid follow-up (sub-)mm observations and for the general study of (sub-)mm transients which typically peak at early times and are short-lived compared to at cm wavelengths (e.g., \cite{laskar2019,ho2019,bright2022,ho2022,andreoni2022}). Supplementary Figure \ref{fig:submm_pred} shows the predicted peak time and flux density of the $230\,\rm{GHz}$ emission from the reverse shock of GRB 221009A which is exceptionally bright at $\approx370\,\rm{mJy}$. Detecting emission on this level is achievable with trivial on-source time using current (sub-)mm facilities but would require a rapid follow-up time on the order of $\approx0.4\,\rm{h}$ post burst to capture this peak. The exceptional predicted (sub-)mm brightness of GRB 221009A is largely due to its close proximity, however such emission could be detectable out to $z\approx10$ (assuming a flat cosmology with $H_{0}=70\,\rm{km}\,\rm{s}^{-1}\,\rm{Mpc}^{-1}$) with the Atacama Large Millimeter/submillimeter Array, although significant cosmological redshift and time dilation become important at such large redshifts where emission observed at $230\,\rm{GHz}$ would correspond to emission emitted at a factor of 10 higher frequency in the rest frame of the emitting region and the peak would be smeared out in time.

\bmhead{Acknowledgments}
%%% Reviewer
We thank the anonymous referee for thoroughly reading an initial draft of this manuscript and for providing insightful comments.
%%% AMI
We thank the staff at the Mullard Radio Astronomy Observatory for the commissioning, maintenance and operation of the Arcminute Microkelvin Imager Large Array, which is supported by the Universities of Cambridge and Oxford. We acknowledge support from the European Research Council under grant ERC-2012-StG-307215 LODESTONE.
%%% ATA
We thank the Hat Creek Radio Observatory staff for carrying out observations with the Allen Telescope Array. The Allen Telescope Array (ATA) refurbishment program and its ongoing operations receive substantial support from Franklin Antonio. Additional contributions from Frank Levinson, Jill Tarter, Jack Welch, the Breakthrough Listen Initiative and other private donors have been instrumental in the renewal of the ATA. Breakthrough Listen is managed by the Breakthrough Initiatives, sponsored by the Breakthrough Prize Foundation. The Paul G. Allen Family Foundation provided major support for the design and construction of the ATA, alongside contributions from Nathan Myhrvold, Xilinx Corporation, Sun Microsystems, and other private donors. The ATA has also been supported by contributions from the US Naval Observatory and the US National Science Foundation.
%%% eMERLIN
\textit{e}-MERLIN is a National Facility operated by the University of Manchester at Jodrell Bank Observatory on behalf of STFC.
%%% SMA
The Submillimeter Array is a joint project between the Smithsonian Astrophysical Observatory and the Academia Sinica Institute of Astronomy and Astrophysics and is funded by the Smithsonian Institution and the Academia Sinica. The authors thank M. Gurwell for reducing the SMA data so quickly and communicating the results which enabled them to be included in this publication. We recognize that Maunakea is a culturally important site for the indigenous Hawaiian people; we are privileged to study the cosmos from its summit.
%%% ASKAP
This scientific work uses data obtained from Inyarrimanha Ilgari Bundara / the Murchison Radio-astronomy Observatory. We acknowledge the Wajarri Yamaji People as the Traditional Owners and native title holders of the Observatory site. The Australian SKA Pathfinder is part of the Australia Telescope National Facility (\url{https://ror.org/05qajvd42}) which is managed by CSIRO. Operation of ASKAP is funded by the Australian Government with support from the National Collaborative Research Infrastructure Strategy. ASKAP uses the resources of the Pawsey Supercomputing Centre. Establishment of ASKAP, the Murchison Radio-astronomy Observatory and the Pawsey Supercomputing Centre are initiatives of the Australian Government, with support from the Government of Western Australia and the Science and Industry Endowment Fund.
%%% CIRADA
This research has made use of the CIRADA cutout service at URL \url{cutouts.cirada.ca}, operated by the Canadian Initiative for Radio Astronomy Data Analysis (CIRADA). CIRADA is funded by a grant from the Canada Foundation for Innovation 2017 Innovation Fund (Project 35999), as well as by the Provinces of Ontario, British Columbia, Alberta, Manitoba and Quebec, in collaboration with the National Research Council of Canada, the US National Radio Astronomy Observatory and Australia’s Commonwealth Scientific and Industrial Research Organisation. 
%%% NRAO
The National Radio Astronomy Observatory is a facility of the National Science Foundation operated under cooperative agreement by Associated Universities, Inc.
%%% Swift
We acknowledge the use of public data from the Swift data archive.
%%% Python
We thank the developers of the Python packages Matplotlib \cite{hunter2007}, NumPy \cite{harris2020}, SciPy \cite{virtanen2020}, and pandas \cite{mckinney2010,pandas2020}.

\bmhead{Data availability statement} Light curve data from the AMI--LA, ATA, ASKAP, \textit{e}-MERLIN, and SMA are given in Extended Data Table \ref{tab:radio_data} and Supplementary Data Table \ref{tab:ami_short}. Full machine readable tables can be found as Supplementary Data files as part of the online material. Continuum images from individual observations are available from the corresponding author upon reasonable request. The results of our fits to the radio light curves are given in Extended Data Table \ref{tab:fitting}.

\bmhead{Author Contributions} JSB and LR wrote the majority of the manuscript. JSB, WF, DRDB, IH, AWP, PHP, SZS, and AS recorded or reduced data from the ATA, or have contributed significantly to the operation of the ATA. JSB, DAG, PSF, DJT recorded or reduced data from the AMI-LA, or have contributed significantly to the operation of the AMI-LA. DW recorded and reduced data from \textit{e}-MERLIN. JKL, GEA, EL, TM, redc. RF, AJvdH, PA, SG provided advice and expertise while observing and interpreting data from GRB221009A.

\bmhead{Competing interests} The authors declare no competing interests.

\newpage

\renewcommand{\tablename}{Extended Data Table}

\begin{table}[h]
\begin{center}
\begin{minipage}{\textwidth}
\centering
\caption{Summary of our radio observations of GRB 221009A with the Arcminute Microkelvin Imager Large Array (AMI--LA), the Allen Telescope Array (ATA), \textit{enhanced}-Multi-Element Radio Linked Interferometer Network (\textit{e}-MERLIN), and Australian Square Kilometre Array Pathfinder (ASKAP). The uncertainties reported here consider only the statistical error on the fit. A 10\%, 10\%, 5\%, and 5\% error should be added in quadrature to the statistical uncertainty for the AMI--LA, ATA, \textit{e}-MERLIN, and ASKAP, respectively, which we account for when plotting or using these data for calculations. We also include public data from the MeerKAT radio telescope and the Japanese VLBI network \cite{laskar_meerkat_gcn,niinuma_jvlbi_gcn}. A full machine-readable table can be found as Supplementary Data files as part of the online material.}\label{tab:radio_data}%
\begin{tabular}{@{}cccccc@{}}
\toprule
Centroid MJD & Flux Density & Flux Density Error & Frequency & Facility\\\newline
 [d] & [mJy] & [mJy] & [GHz] &\\
\midrule
59861.7668&39.43&0.18&15.5&AMI\textsuperscript{1} \\
59861.9414&9.39&0.39&6.0&ATA\textsuperscript{2} \\ 
59862.0456&3.97&0.5&3.0&ATA \\
59862.1242&37.64&0.91&10.0&ATA \\ 
59862.1242&28.39&0.59&8.0&ATA \\
... & ... & ... & ... & ...\\
\botrule
\end{tabular}
\footnotetext[1]{Arcminute Microkelvin Imager Large Array \cite{zwart2008,hickish2018}.}
\footnotetext[2]{Allen Telescope Array.}
\end{minipage}
\end{center}
\end{table}

\newpage

\begin{table}[h]
\begin{center}
\begin{minipage}{\textwidth}
\centering
\caption{Best fit parameters for our smoothly broken power law fits to the radio light curves of GRB 221009A between $3\,\rm{GHz}$ and $17.69\,\rm{GHz}$. A smoothing parameter $s=2$ is used. For further information on the fitting process see the Methods. The time of the light curve peak and the flux density at the peak are given by $t_{b}$ and $A$, respectively. Fits are not made to the 10 and 1.5\,GHz data due to lack of a clear peak and insufficient data, respectively.}\label{tab:fitting}%
\begin{tabular}{@{}ccc@{}}
\toprule
Frequency (GHz) & $A$ (mJy) & $t_{b}$  (hours)\\
\midrule
%1.3 & $2.7\pm3.9$ & $2.2\pm1.0$ \\
3 & $13.8\pm0.6$ & $40.4\pm4.9$ \\
5 & $19.8\pm1.2$ & $23.8\pm1.9$ \\
8 & $29.3\pm2.3$ & $14.8\pm1.6$ \\
%10 & $38.1\pm18.1$ & $0.46\pm0.26$ \\
%14.25 & $44.7\pm1.7$ & $0.3275\pm0.0099$ \\
%16.75 & $54.7\pm1.3$ & $0.2699\pm0.0050$ \\
13.31 & $43.6\pm0.5 $ & $9.2\pm0.1$ \\
13.94 & $44.6\pm0.5$ & $8.7\pm0.1$\\ 
14.56 & $46.5\pm0.5$ & $8.29\pm0.09$\\ 
15.19 & $48.6\pm0.5$ & $7.86\pm0.09$\\ 
15.81 & $50.6\pm0.5$ & $7.39\pm0.09$\\ 
16.44 & $53.0\pm0.6$ & $6.95\pm0.09$\\ 
17.06 & $55.9\pm0.6$ & $6.55\pm0.08$\\ 
17.69 & $57.2\pm0.6$ & $6.09\pm0.09$\\ 
\botrule
\end{tabular}
\end{minipage}
\end{center}
\end{table}

\newpage

\newpage

\newpage

\newpage

\renewcommand{\theequation}{S\arabic{equation}}
\renewcommand{\figurename}{Extended Data Figure}
\setcounter{figure}{0}
\setcounter{section}{0}

\newpage

\begin{figure}[H]
\centering
\includegraphics[width=0.9\textwidth]{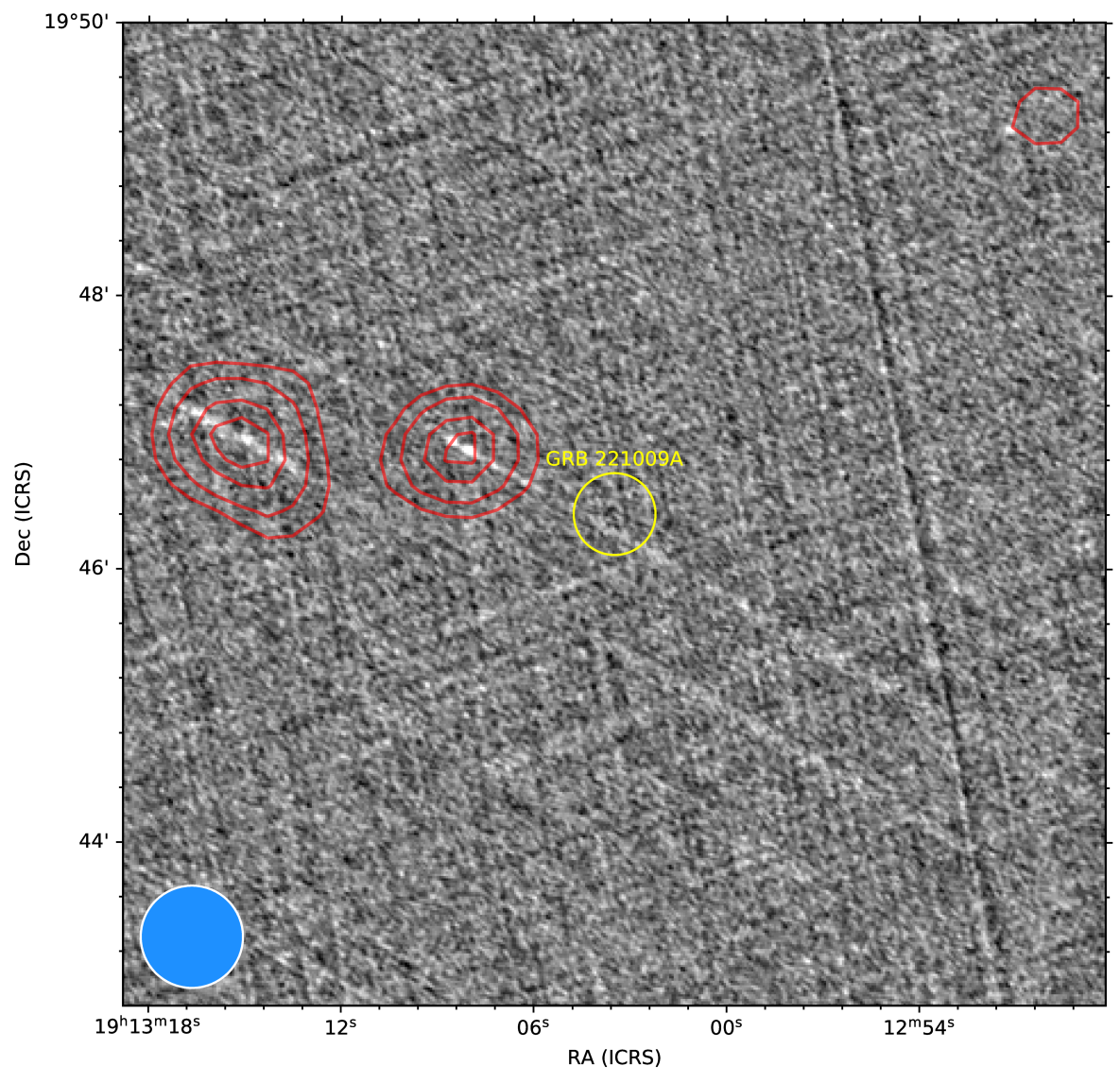}
\caption{\textbf{Very Large Array Sky Survey archival observations of the field of GRB 221009A.} The Very Large Array Sky Survey (VLASS; version 2.2; \cite{lacy2020}) observation of the field of GRB 221009A, with National Radio Astronomy Observatory Very Large Array Sky Survey (NVSS; \cite{condon1998}) contours over-plotted in red. The restoring beam for the NVSS image is shown as a blue circle in the bottom left of the image, the restoring beam for VLASS is significantly smaller and is not shown, but has a major and minor axis of $3.31''$ and $2.29''$, respectively, at a position angle of $51.04^{\circ}$. The yellow circle is centred on the position of GRB 221009A \cite{atri_gcn} and has a radius of $18''$. No significant emission from either survey is seen at the position of GRB 221009A. The most constraining limit is from VLASS for which we measure a root mean square three sigma upper limit of $\sim450\,\upmu\rm{Jy/beam}$. A number of deconvolution/calibration artefacts are present in the quick-look VLASS image and likely are the result of incomplete deconvolution of bright sources. These manifest as `streaks' most notable between North and South on the East side of the image.
}\label{fig:archival_radio}
\end{figure}

\newpage

\begin{figure}[H]
\centering
\includegraphics[width=0.9\textwidth]{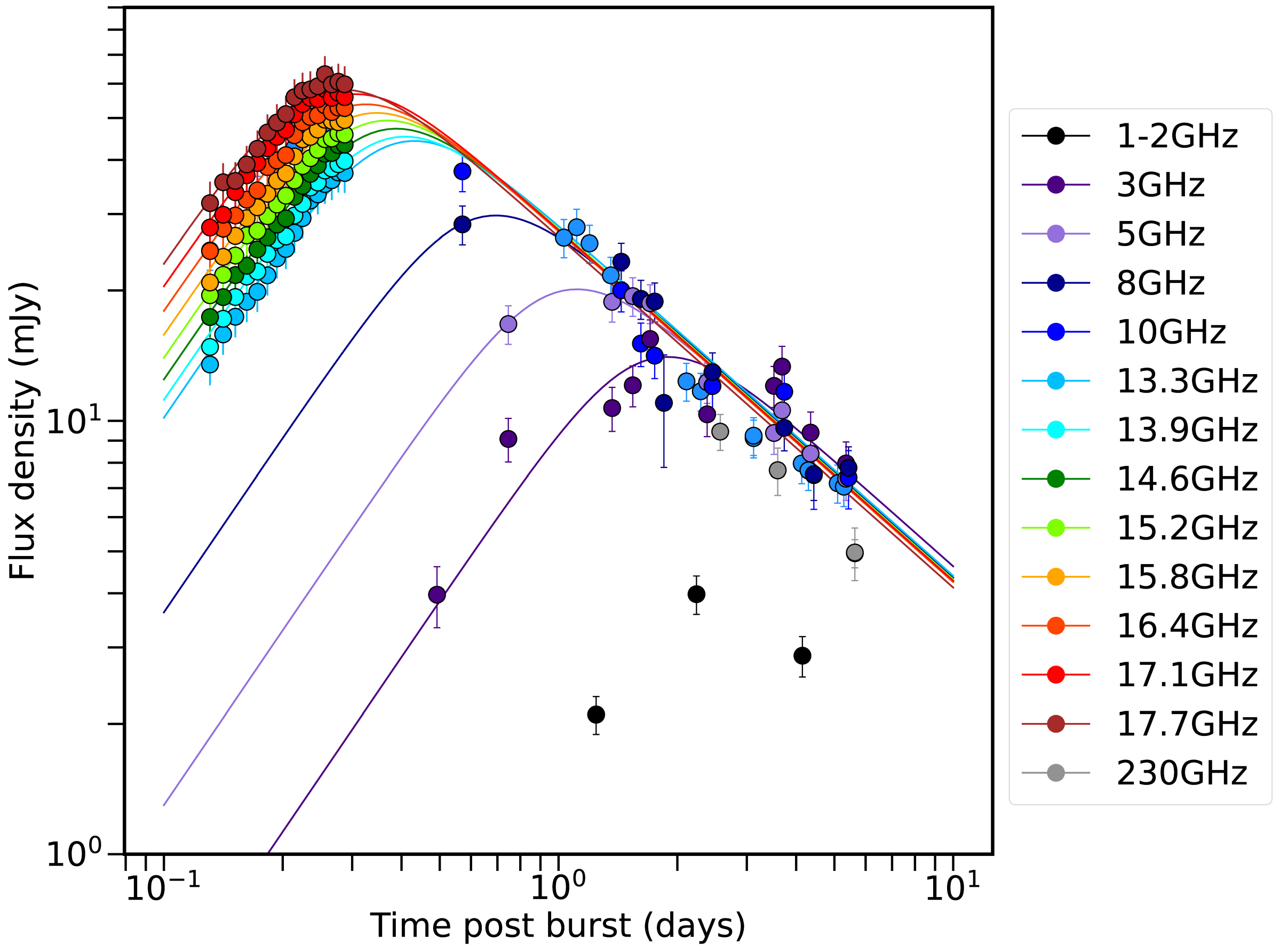}
\caption{\textbf{Power-law fits to our multi-frequency radio observations of GRB221009A.} A broken power-law fit to each radio band where we observe a clear peak. The results of the fits are giving in Extended Data Table \ref{tab:fitting}. The rise and decay power law indices follow $F_{rise}\propto t^{1.34\pm0.02}$ and $F_{decay}\propto t^{-0.82\pm0.04}$, respectively. Error bars represent $1\sigma$ uncertainties.}
\label{fig:fits}
\end{figure}

\newpage

%\begin{appendices}

%\section{An Appendix}\label{sec:appendix}




%\end{appendices}


\bibliography{GRB221009A}% common bib file

\begin{thebibliography}{10}
\expandafter\ifx\csname url\endcsname\relax
  \def\url#1{\burl{#1}}\fi
\expandafter\ifx\csname urlprefix\endcsname\relax\def\urlprefix{URL }\fi
\providecommand{\bibinfo}[2]{#2}
\providecommand{\eprint}[2][]{\url{#2}}
\providecommand{\doi}[1]{\url{https://doi.org/#1}}
\bibcommenthead

\bibitem{kouveliotou1993}
\bibinfo{author}{{Kouveliotou}, C.} \emph{et~al.}
\newblock \bibinfo{title}{{Identification of Two Classes of Gamma-Ray Bursts}}.
\newblock \emph{\bibinfo{journal}{\apjl}} \textbf{\bibinfo{volume}{413}}, \bibinfo{pages}{L101} (\bibinfo{year}{1993}).

\bibitem{galama1998}
\bibinfo{author}{{Galama}, T.~J.} \emph{et~al.}
\newblock \bibinfo{title}{{An unusual supernova in the error box of the {\ensuremath{\gamma}}-ray burst of 25 April 1998}}.
\newblock \emph{\bibinfo{journal}{\nat}} \textbf{\bibinfo{volume}{395}}, \bibinfo{pages}{670--672} (\bibinfo{year}{1998}).

\bibitem{levan2016}
\bibinfo{author}{{Levan}, A.} \emph{et~al.}
\newblock \bibinfo{title}{{Gamma-Ray Burst Progenitors}}.
\newblock \emph{\bibinfo{journal}{\ssr}} \textbf{\bibinfo{volume}{202}}, \bibinfo{pages}{33--78} (\bibinfo{year}{2016}).

\bibitem{rees1994}
\bibinfo{author}{{Rees}, M.~J.} \& \bibinfo{author}{{Meszaros}, P.}
\newblock \bibinfo{title}{{Unsteady Outflow Models for Cosmological Gamma-Ray Bursts}}.
\newblock \emph{\bibinfo{journal}{\apjl}} \textbf{\bibinfo{volume}{430}}, \bibinfo{pages}{L93} (\bibinfo{year}{1994}).

\bibitem{kobayashi1997}
\bibinfo{author}{{Kobayashi}, S.}, \bibinfo{author}{{Piran}, T.} \& \bibinfo{author}{{Sari}, R.}
\newblock \bibinfo{title}{{Can Internal Shocks Produce the Variability in Gamma-Ray Bursts?}}
\newblock \emph{\bibinfo{journal}{\apj}} \textbf{\bibinfo{volume}{490}}, \bibinfo{pages}{92} (\bibinfo{year}{1997}).

\bibitem{granot2011}
\bibinfo{author}{{Granot}, J.}, \bibinfo{author}{{Komissarov}, S.~S.} \& \bibinfo{author}{{Spitkovsky}, A.}
\newblock \bibinfo{title}{{Impulsive acceleration of strongly magnetized relativistic flows}}.
\newblock \emph{\bibinfo{journal}{\mnras}} \textbf{\bibinfo{volume}{411}}, \bibinfo{pages}{1323--1353} (\bibinfo{year}{2011}).

\bibitem{perley2014}
\bibinfo{author}{{Perley}, D.~A.} \emph{et~al.}
\newblock \bibinfo{title}{{The Afterglow of GRB 130427A from 1 to {}10$^{16}$ GHz}}.
\newblock \emph{\bibinfo{journal}{\apj}} \textbf{\bibinfo{volume}{781}}, \bibinfo{pages}{37} (\bibinfo{year}{2014}).

\bibitem{vanderhorst2014}
\bibinfo{author}{{van der Horst}, A.~J.} \emph{et~al.}
\newblock \bibinfo{title}{{A comprehensive radio view of the extremely bright gamma-ray burst 130427A}}.
\newblock \emph{\bibinfo{journal}{\mnras}} \textbf{\bibinfo{volume}{444}}, \bibinfo{pages}{3151--3163} (\bibinfo{year}{2014}).

\bibitem{laskar2013}
\bibinfo{author}{{Laskar}, T.} \emph{et~al.}
\newblock \bibinfo{title}{{A Reverse Shock in GRB 130427A}}.
\newblock \emph{\bibinfo{journal}{\apj}} \textbf{\bibinfo{volume}{776}}, \bibinfo{pages}{119} (\bibinfo{year}{2013}).

\bibitem{piran1999}
\bibinfo{author}{{Piran}, T.}
\newblock \bibinfo{title}{{Gamma-ray bursts and the fireball model}}.
\newblock \emph{\bibinfo{journal}{\physrep}} \textbf{\bibinfo{volume}{314}}, \bibinfo{pages}{575--667} (\bibinfo{year}{1999}).

\bibitem{granot2002}
\bibinfo{author}{{Granot}, J.} \& \bibinfo{author}{{Sari}, R.}
\newblock \bibinfo{title}{{The Shape of Spectral Breaks in Gamma-Ray Burst Afterglows}}.
\newblock \emph{\bibinfo{journal}{\apj}} \textbf{\bibinfo{volume}{568}}, \bibinfo{pages}{820--829} (\bibinfo{year}{2002}).

\bibitem{ressler2017}
\bibinfo{author}{{Ressler}, S.~M.} \& \bibinfo{author}{{Laskar}, T.}
\newblock \bibinfo{title}{{Thermal Electrons in Gamma-Ray Burst Afterglows}}.
\newblock \emph{\bibinfo{journal}{\apj}} \textbf{\bibinfo{volume}{845}}, \bibinfo{pages}{150} (\bibinfo{year}{2017}).

\bibitem{cucchiara2011}
\bibinfo{author}{{Cucchiara}, A.} \emph{et~al.}
\newblock \bibinfo{title}{{A Photometric Redshift of $z\sim9.4$ for GRB 090429B}}.
\newblock \emph{\bibinfo{journal}{\apj}} \textbf{\bibinfo{volume}{736}}, \bibinfo{pages}{7} (\bibinfo{year}{2011}).

\bibitem{fynbo2009}
\bibinfo{author}{{Fynbo}, J.~P.~U.} \emph{et~al.}
\newblock \bibinfo{title}{{Low-resolution Spectroscopy of Gamma-ray Burst Optical Afterglows: Biases in the Swift Sample and Characterization of the Absorbers}}.
\newblock \emph{\bibinfo{journal}{\apjs}} \textbf{\bibinfo{volume}{185}}, \bibinfo{pages}{526--573} (\bibinfo{year}{2009}).

\bibitem{robsertson2012}
\bibinfo{author}{{Robertson}, B.~E.} \& \bibinfo{author}{{Ellis}, R.~S.}
\newblock \bibinfo{title}{{Connecting the Gamma Ray Burst Rate and the Cosmic Star Formation History: Implications for Reionization and Galaxy Evolution}}.
\newblock \emph{\bibinfo{journal}{\apj}} \textbf{\bibinfo{volume}{744}}, \bibinfo{pages}{95} (\bibinfo{year}{2012}).

\bibitem{kruhler2015}
\bibinfo{author}{{Kr{\"u}hler}, T.} \emph{et~al.}
\newblock \bibinfo{title}{{GRB hosts through cosmic time. VLT/X-Shooter emission-line spectroscopy of 96 {\ensuremath{\gamma}}-ray-burst-selected galaxies at $0.1 <z < 3.6$}}.
\newblock \emph{\bibinfo{journal}{\aap}} \textbf{\bibinfo{volume}{581}}, \bibinfo{pages}{A125} (\bibinfo{year}{2015}).

\bibitem{pescalli2016}
\bibinfo{author}{{Pescalli}, A.} \emph{et~al.}
\newblock \bibinfo{title}{{The rate and luminosity function of long gamma ray bursts}}.
\newblock \emph{\bibinfo{journal}{\aap}} \textbf{\bibinfo{volume}{587}}, \bibinfo{pages}{A40} (\bibinfo{year}{2016}).

\bibitem{matthews2021}
\bibinfo{author}{{Matthews}, A.~M.}, \bibinfo{author}{{Condon}, J.~J.}, \bibinfo{author}{{Cotton}, W.~D.} \& \bibinfo{author}{{Mauch}, T.}
\newblock \bibinfo{title}{{Cosmic Star Formation History Measured at 1.4 GHz}}.
\newblock \emph{\bibinfo{journal}{\apj}} \textbf{\bibinfo{volume}{914}}, \bibinfo{pages}{126} (\bibinfo{year}{2021}).

\bibitem{soderberg2006}
\bibinfo{author}{{Soderberg}, A.~M.} \emph{et~al.}
\newblock \bibinfo{title}{{Relativistic ejecta from X-ray flash XRF 060218 and the rate of cosmic explosions}}.
\newblock \emph{\bibinfo{journal}{\nat}} \textbf{\bibinfo{volume}{442}}, \bibinfo{pages}{1014--1017} (\bibinfo{year}{2006}).

\bibitem{stanek2006}
\bibinfo{author}{{Stanek}, K.~Z.} \emph{et~al.}
\newblock \bibinfo{title}{{Protecting Life in the Milky Way: Metals Keep the GRBs Away}}.
\newblock \emph{\bibinfo{journal}{\actaa}} \textbf{\bibinfo{volume}{56}}, \bibinfo{pages}{333--345} (\bibinfo{year}{2006}).

\bibitem{guetta2007}
\bibinfo{author}{{Guetta}, D.} \& \bibinfo{author}{{Della Valle}, M.}
\newblock \bibinfo{title}{{On the Rates of Gamma-Ray Bursts and Type Ib/c Supernovae}}.
\newblock \emph{\bibinfo{journal}{\apjl}} \textbf{\bibinfo{volume}{657}}, \bibinfo{pages}{L73--L76} (\bibinfo{year}{2007}).

\bibitem{vanderhorst2008}
\bibinfo{author}{{van der Horst}, A.~J.} \emph{et~al.}
\newblock \bibinfo{title}{{Detailed study of the GRB 030329 radio afterglow deep into the non-relativistic phase}}.
\newblock \emph{\bibinfo{journal}{\aap}} \textbf{\bibinfo{volume}{480}}, \bibinfo{pages}{35--43} (\bibinfo{year}{2008}).

\bibitem{anderson2014}
\bibinfo{author}{{Anderson}, G.~E.} \emph{et~al.}
\newblock \bibinfo{title}{{Probing the bright radio flare and afterglow of GRB 130427A with the Arcminute Microkelvin Imager}}.
\newblock \emph{\bibinfo{journal}{\mnras}} \textbf{\bibinfo{volume}{440}}, \bibinfo{pages}{2059--2065} (\bibinfo{year}{2014}).

\bibitem{bright2019}
\bibinfo{author}{{Bright}, J.~S.} \emph{et~al.}
\newblock \bibinfo{title}{{A detailed radio study of the energetic, nearby, and puzzling GRB 171010A}}.
\newblock \emph{\bibinfo{journal}{\mnras}} \textbf{\bibinfo{volume}{486}}, \bibinfo{pages}{2721--2729} (\bibinfo{year}{2019}).

\bibitem{meegan2009}
\bibinfo{author}{{Meegan}, C.} \emph{et~al.}
\newblock \bibinfo{title}{{The Fermi Gamma-ray Burst Monitor}}.
\newblock \emph{\bibinfo{journal}{\apj}} \textbf{\bibinfo{volume}{702}}, \bibinfo{pages}{791--804} (\bibinfo{year}{2009}).

\bibitem{veres2022}
\bibinfo{author}{{Veres}, P.} \emph{et~al.}
\newblock \bibinfo{title}{{GRB 221009A: Fermi GBM detection of an extraordinarily bright GRB}}.
\newblock \emph{\bibinfo{journal}{GRB Coordinates Network}} \textbf{\bibinfo{volume}{32636}} (\bibinfo{year}{2022}).

\bibitem{dichiara2022}
\bibinfo{author}{{Dichiara}, S.} \emph{et~al.}
\newblock \bibinfo{title}{{Swift J1913.1+1946 a new bright hard X-ray and optical transient}}.
\newblock \emph{\bibinfo{journal}{GRB Coordinates Network}} \textbf{\bibinfo{volume}{32632}} (\bibinfo{year}{2022}).

\bibitem{deugartepostigo2022}
\bibinfo{author}{{de Ugarte Postigo}, A.} \emph{et~al.}
\newblock \bibinfo{title}{{GRB 221009A: Redshift from X-shooter/VLT}}.
\newblock \emph{\bibinfo{journal}{GRB Coordinates Network}} \textbf{\bibinfo{volume}{32648}} (\bibinfo{year}{2022}).

\bibitem{castrotirado2022}
\bibinfo{author}{{Castro-Tirado}, A.~J.} \emph{et~al.}
\newblock \bibinfo{title}{{GRB 221009A: 10.4m GTC spectroscopic redshift confirmation}}.
\newblock \emph{\bibinfo{journal}{GRB Coordinates Network}} \textbf{\bibinfo{volume}{32686}} (\bibinfo{year}{2022}).

\bibitem{frederiks2022}
\bibinfo{author}{{Frederiks}, D.} \emph{et~al.}
\newblock \bibinfo{title}{{Konus-Wind detection of GRB 221009A}}.
\newblock \emph{\bibinfo{journal}{GRB Coordinates Network}} \textbf{\bibinfo{volume}{32668}} (\bibinfo{year}{2022}).

\bibitem{williams2023}
\bibinfo{author}{{Williams}, M.~A.} \emph{et~al.}
\newblock \bibinfo{title}{{GRB 221009A: Discovery of an Exceptionally Rare Nearby and Energetic Gamma-Ray Burst}}.
\newblock \emph{\bibinfo{journal}{arXiv e-prints}} \bibinfo{pages}{arXiv:2302.03642} (\bibinfo{year}{2023}).

\bibitem{laskar2023}
\bibinfo{author}{{Laskar}, T.} \emph{et~al.}
\newblock \bibinfo{title}{{The Radio to GeV Afterglow of GRB 221009A}}.
\newblock \emph{\bibinfo{journal}{arXiv e-prints}} \bibinfo{pages}{arXiv:2302.04388} (\bibinfo{year}{2023}).

\bibitem{oconnor2023}
\bibinfo{author}{{O'Connor}, B.} \emph{et~al.}
\newblock \bibinfo{title}{{A structured jet explains the extreme GRB 221009A}}.
\newblock \emph{\bibinfo{journal}{arXiv e-prints}} \bibinfo{pages}{arXiv:2302.07906} (\bibinfo{year}{2023}).

\bibitem{2022ATel15664....1I}
\bibinfo{author}{{Iwakiri}, W.} \emph{et~al.}
\newblock \bibinfo{title}{{GRB 221009A: NICER follow-up observations}}.
\newblock \emph{\bibinfo{journal}{The Astronomer's Telegram}} \textbf{\bibinfo{volume}{15664}} (\bibinfo{year}{2022}).

\bibitem{2022GCN.32645....1B}
\bibinfo{author}{{Belkin}, S.}, \bibinfo{author}{{Pozanenko}, A.}, \bibinfo{author}{{Klunko}, E.}, \bibinfo{author}{{Pankov}, N.} \& \bibinfo{author}{{GRB IKI FuN}}.
\newblock \bibinfo{title}{{GRB 221009A (Swift J1913.1+1946): Mondy optical observations}}.
\newblock \emph{\bibinfo{journal}{GRB Coordinates Network}} \textbf{\bibinfo{volume}{32645}} (\bibinfo{year}{2022}).

\bibitem{2022GCN.32646....1D}
\bibinfo{author}{{de Wet}, S.}, \bibinfo{author}{{Groot}, P.~J.} \& \bibinfo{author}{{Meerlicht Consortium}}.
\newblock \bibinfo{title}{{GRB 221009A (Swift J1913.1+1946): MeerLICHT observations}}.
\newblock \emph{\bibinfo{journal}{GRB Coordinates Network}} \textbf{\bibinfo{volume}{32646}} (\bibinfo{year}{2022}).

\bibitem{2022GCN.32652....1B}
\bibinfo{author}{{Brivio}, R.} \emph{et~al.}
\newblock \bibinfo{title}{{GRB 221009A: REM optical and NIR detection of the afterglow}}.
\newblock \emph{\bibinfo{journal}{GRB Coordinates Network}} \textbf{\bibinfo{volume}{32652}} (\bibinfo{year}{2022}).

\bibitem{2022GCN.32659....1P}
\bibinfo{author}{{Paek}, G. S.~H.}, \bibinfo{author}{{Im}, M.}, \bibinfo{author}{{Urata}, Y.} \& \bibinfo{author}{{Sung}, H.-I.}
\newblock \bibinfo{title}{{GRB 221009A: Multi-color detection of the optical}}.
\newblock \emph{\bibinfo{journal}{GRB Coordinates Network}} \textbf{\bibinfo{volume}{32659}} (\bibinfo{year}{2022}).

\bibitem{2022GCN.32669....1V}
\bibinfo{author}{{Vidal}, E.}, \bibinfo{author}{{Zheng}, W.}, \bibinfo{author}{{Filippenko}, A.~V.} \& \bibinfo{author}{{KAIT GRB team}}.
\newblock \bibinfo{title}{{GRB 221009A/Swift J1913.1+1946: Lick/Nickel telescope optical observations}}.
\newblock \emph{\bibinfo{journal}{GRB Coordinates Network}} \textbf{\bibinfo{volume}{32669}} (\bibinfo{year}{2022}).

\bibitem{2022GCN.32676....1D}
\bibinfo{author}{{de Ugarte Postigo}, A.} \emph{et~al.}
\newblock \bibinfo{title}{{GRB 221009A: NOEMA mm detection}}.
\newblock \emph{\bibinfo{journal}{GRB Coordinates Network}} \textbf{\bibinfo{volume}{32676}} (\bibinfo{year}{2022}).

\bibitem{2022GCN.32678....1G}
\bibinfo{author}{{Groot}, P.~J.} \emph{et~al.}
\newblock \bibinfo{title}{{GRB 221009A: BlackGEM optical observations}}.
\newblock \emph{\bibinfo{journal}{GRB Coordinates Network}} \textbf{\bibinfo{volume}{32678}} (\bibinfo{year}{2022}).

\bibitem{zwart2008}
\bibinfo{author}{{Zwart}, J.~T.~L.} \emph{et~al.}
\newblock \bibinfo{title}{{The Arcminute Microkelvin Imager}}.
\newblock \emph{\bibinfo{journal}{\mnras}} \textbf{\bibinfo{volume}{391}}, \bibinfo{pages}{1545--1558} (\bibinfo{year}{2008}).

\bibitem{hickish2018}
\bibinfo{author}{{Hickish}, J.} \emph{et~al.}
\newblock \bibinfo{title}{{A digital correlator upgrade for the Arcminute MicroKelvin Imager}}.
\newblock \emph{\bibinfo{journal}{\mnras}} \textbf{\bibinfo{volume}{475}}, \bibinfo{pages}{5677--5687} (\bibinfo{year}{2018}).

\bibitem{levan2023}
\bibinfo{author}{{Levan}, A.~J.} \emph{et~al.}
\newblock \bibinfo{title}{{The first JWST spectrum of a GRB afterglow: No bright supernova in observations of the brightest GRB of all time, GRB 221009A}}.
\newblock \emph{\bibinfo{journal}{arXiv e-prints}} \bibinfo{pages}{arXiv:2302.07761} (\bibinfo{year}{2023}).

\bibitem{taylor2004}
\bibinfo{author}{{Taylor}, G.~B.}, \bibinfo{author}{{Frail}, D.~A.}, \bibinfo{author}{{Berger}, E.} \& \bibinfo{author}{{Kulkarni}, S.~R.}
\newblock \bibinfo{title}{{The Angular Size and Proper Motion of the Afterglow of GRB 030329}}.
\newblock \emph{\bibinfo{journal}{\apjl}} \textbf{\bibinfo{volume}{609}}, \bibinfo{pages}{L1--L4} (\bibinfo{year}{2004}).

\bibitem{frail2005}
\bibinfo{author}{{Frail}, D.~A.} \emph{et~al.}
\newblock \bibinfo{title}{{Accurate Calorimetry of GRB 030329}}.
\newblock \emph{\bibinfo{journal}{\apj}} \textbf{\bibinfo{volume}{619}}, \bibinfo{pages}{994--998} (\bibinfo{year}{2005}).

\bibitem{alexander2019}
\bibinfo{author}{{Alexander}, K.~D.} \emph{et~al.}
\newblock \bibinfo{title}{{An Unexpectedly Small Emission Region Size Inferred from Strong High-frequency Diffractive Scintillation in GRB 161219B}}.
\newblock \emph{\bibinfo{journal}{\apj}} \textbf{\bibinfo{volume}{870}}, \bibinfo{pages}{67} (\bibinfo{year}{2019}).

\bibitem{anderson2022}
\bibinfo{author}{{Anderson}, G.~E.} \emph{et~al.}
\newblock \bibinfo{title}{{Rapid radio brightening of GRB 210702A}}.
\newblock \emph{\bibinfo{journal}{arXiv e-prints}} \bibinfo{pages}{arXiv:2211.11212} (\bibinfo{year}{2022}).

\bibitem{liang2010}
\bibinfo{author}{{Liang}, E.-W.} \emph{et~al.}
\newblock \bibinfo{title}{{Constraining Gamma-ray Burst Initial Lorentz Factor with the Afterglow Onset Feature and Discovery of a Tight {\ensuremath{\Gamma}}$_{0}$-E$_{{\ensuremath{\gamma}},iso}$ Correlation}}.
\newblock \emph{\bibinfo{journal}{\apj}} \textbf{\bibinfo{volume}{725}}, \bibinfo{pages}{2209--2224} (\bibinfo{year}{2010}).

\bibitem{liang2015}
\bibinfo{author}{{Liang}, E.-W.} \emph{et~al.}
\newblock \bibinfo{title}{{A Tight L$_{iso}$ - E$_{p,z}$ - $\Gamma_{0}$ Correlation of Gamma-Ray Bursts}}.
\newblock \emph{\bibinfo{journal}{\apj}} \textbf{\bibinfo{volume}{813}}, \bibinfo{pages}{116} (\bibinfo{year}{2015}).

\bibitem{anderson2018}
\bibinfo{author}{{Anderson}, G.~E.} \emph{et~al.}
\newblock \bibinfo{title}{{The Arcminute Microkelvin Imager catalogue of gamma-ray burst afterglows at 15.7 GHz}}.
\newblock \emph{\bibinfo{journal}{\mnras}} \textbf{\bibinfo{volume}{473}}, \bibinfo{pages}{1512--1536} (\bibinfo{year}{2018}).

\bibitem{sari1998}
\bibinfo{author}{{Sari}, R.}, \bibinfo{author}{{Piran}, T.} \& \bibinfo{author}{{Narayan}, R.}
\newblock \bibinfo{title}{{Spectra and Light Curves of Gamma-Ray Burst Afterglows}}.
\newblock \emph{\bibinfo{journal}{\apjl}} \textbf{\bibinfo{volume}{497}}, \bibinfo{pages}{L17--L20} (\bibinfo{year}{1998}).

\bibitem{chevalier2000}
\bibinfo{author}{{Chevalier}, R.~A.} \& \bibinfo{author}{{Li}, Z.-Y.}
\newblock \bibinfo{title}{{Wind Interaction Models for Gamma-Ray Burst Afterglows: The Case for Two Types of Progenitors}}.
\newblock \emph{\bibinfo{journal}{\apj}} \textbf{\bibinfo{volume}{536}}, \bibinfo{pages}{195--212} (\bibinfo{year}{2000}).

\bibitem{yost2003}
\bibinfo{author}{{Yost}, S.~A.}, \bibinfo{author}{{Harrison}, F.~A.}, \bibinfo{author}{{Sari}, R.} \& \bibinfo{author}{{Frail}, D.~A.}
\newblock \bibinfo{title}{{A Study of the Afterglows of Four Gamma-Ray Bursts: Constraining the Explosion and Fireball Model}}.
\newblock \emph{\bibinfo{journal}{\apj}} \textbf{\bibinfo{volume}{597}}, \bibinfo{pages}{459--473} (\bibinfo{year}{2003}).

\bibitem{aksulu2022}
\bibinfo{author}{{Aksulu}, M.~D.}, \bibinfo{author}{{Wijers}, R.~A.~M.~J.}, \bibinfo{author}{{van Eerten}, H.~J.} \& \bibinfo{author}{{van der Horst}, A.~J.}
\newblock \bibinfo{title}{{Exploring the GRB population: robust afterglow modelling}}.
\newblock \emph{\bibinfo{journal}{\mnras}} \textbf{\bibinfo{volume}{511}}, \bibinfo{pages}{2848--2867} (\bibinfo{year}{2022}).

\bibitem{rees1998}
\bibinfo{author}{{Rees}, M.~J.} \& \bibinfo{author}{{M{\'e}sz{\'a}ros}, P.}
\newblock \bibinfo{title}{{Refreshed Shocks and Afterglow Longevity in Gamma-Ray Bursts}}.
\newblock \emph{\bibinfo{journal}{\apjl}} \textbf{\bibinfo{volume}{496}}, \bibinfo{pages}{L1--L4} (\bibinfo{year}{1998}).

\bibitem{mcmullin2007}
\bibinfo{author}{{McMullin}, J.~P.}, \bibinfo{author}{{Waters}, B.}, \bibinfo{author}{{Schiebel}, D.}, \bibinfo{author}{{Young}, W.} \& \bibinfo{author}{{Golap}, K.}
\newblock \bibinfo{editor}{{Shaw}, R.~A.}, \bibinfo{editor}{{Hill}, F.} \& \bibinfo{editor}{{Bell}, D.~J.} (eds) \emph{\bibinfo{title}{{CASA Architecture and Applications}}}.
\newblock (eds \bibinfo{editor}{{Shaw}, R.~A.}, \bibinfo{editor}{{Hill}, F.} \& \bibinfo{editor}{{Bell}, D.~J.}) \emph{\bibinfo{booktitle}{Astronomical Data Analysis Software and Systems XVI}}, Vol. \bibinfo{volume}{376} of \emph{\bibinfo{series}{Astronomical Society of the Pacific Conference Series}}, \bibinfo{pages}{127} (\bibinfo{year}{2007}).

\bibitem{thecasateam2022}
\bibinfo{author}{{THE CASA TEAM}} \emph{et~al.}
\newblock \bibinfo{title}{{CASA, the Common Astronomy Software Applications for Radio Astronomy}}.
\newblock \emph{\bibinfo{journal}{arXiv e-prints}} \bibinfo{pages}{arXiv:2210.02276} (\bibinfo{year}{2022}).

\bibitem{welch2017}
\bibinfo{author}{{Welch}, W.~J.} \emph{et~al.}
\newblock \bibinfo{title}{{New Cooled Feeds for the Allen Telescope Array}}.
\newblock \emph{\bibinfo{journal}{\pasp}} \textbf{\bibinfo{volume}{129}}, \bibinfo{pages}{045002} (\bibinfo{year}{2017}).

\bibitem{xGPU}
\bibinfo{author}{{Clark}, M.~A.}, \bibinfo{author}{{LaPlante}, P.~C.} \& \bibinfo{author}{{Greenhill}, L.~J.}
\newblock \bibinfo{title}{{Accelerating radio astronomy cross-correlation with graphics processing units}}.
\newblock \emph{\bibinfo{journal}{International Journal of High Performance Computing Applications}} \textbf{\bibinfo{volume}{27}}, \bibinfo{pages}{178--192} (\bibinfo{year}{2013}).

\bibitem{aoflagger}
\bibinfo{author}{{Offringa}, A.~R.}
\newblock \bibinfo{title}{{AOFlagger: RFI Software}}.
\newblock \bibinfo{howpublished}{Astrophysics Source Code Library, record ascl:1010.017} (\bibinfo{year}{2010}).
\newblock \eprint{1010.017}.

\bibitem{briggsphd}
\bibinfo{author}{{Briggs}, D.~S.}
\newblock \emph{\bibinfo{title}{{High fidelity deconvolution of moderately resolved sources}}}.
\newblock Ph.D. thesis, \bibinfo{school}{New Mexico Institute of Mining and Technology} (\bibinfo{year}{1995}).

\bibitem{moldon2021}
\bibinfo{author}{{Moldon}, J.}
\newblock \bibinfo{title}{{eMCP: e-MERLIN CASA pipeline}}.
\newblock \bibinfo{howpublished}{Astrophysics Source Code Library, record ascl:2109.006} (\bibinfo{year}{2021}).
\newblock \eprint{2109.006}.

\bibitem{johnston2007}
\bibinfo{author}{{Johnston}, S.} \emph{et~al.}
\newblock \bibinfo{title}{{Science with the Australian Square Kilometre Array Pathfinder}}.
\newblock \emph{\bibinfo{journal}{\pasa}} \textbf{\bibinfo{volume}{24}}, \bibinfo{pages}{174--188} (\bibinfo{year}{2007}).

\bibitem{hotan2021}
\bibinfo{author}{{Hotan}, A.~W.} \emph{et~al.}
\newblock \bibinfo{title}{{Australian square kilometre array pathfinder: I. system description}}.
\newblock \emph{\bibinfo{journal}{\pasa}} \textbf{\bibinfo{volume}{38}}, \bibinfo{pages}{e009} (\bibinfo{year}{2021}).

\bibitem{guzman2019}
\bibinfo{author}{{Guzman}, J.} \emph{et~al.}
\newblock \bibinfo{title}{{ASKAPsoft: ASKAP science data processor software}}.
\newblock \bibinfo{howpublished}{Astrophysics Source Code Library, record ascl:1912.003} (\bibinfo{year}{2019}).
\newblock \eprint{1912.003}.

\bibitem{hale2021}
\bibinfo{author}{{Hale}, C.~L.} \emph{et~al.}
\newblock \bibinfo{title}{{The Rapid ASKAP Continuum Survey Paper II: First Stokes I Source Catalogue Data Release}}.
\newblock \emph{\bibinfo{journal}{\pasa}} \textbf{\bibinfo{volume}{38}}, \bibinfo{pages}{e058} (\bibinfo{year}{2021}).

\bibitem{zhang2004}
\bibinfo{author}{{Zhang}, B.} \& \bibinfo{author}{{M{\'e}sz{\'a}ros}, P.}
\newblock \bibinfo{title}{{Gamma-Ray Bursts: progress, problems \& prospects}}.
\newblock \emph{\bibinfo{journal}{International Journal of Modern Physics A}} \textbf{\bibinfo{volume}{19}}, \bibinfo{pages}{2385--2472} (\bibinfo{year}{2004}).

\bibitem{rickett1990}
\bibinfo{author}{{Rickett}, B.~J.}
\newblock \bibinfo{title}{{Radio propagation through the turbulent interstellar plasma.}}
\newblock \emph{\bibinfo{journal}{\araa}} \textbf{\bibinfo{volume}{28}}, \bibinfo{pages}{561--605} (\bibinfo{year}{1990}).

\bibitem{cordes2002}
\bibinfo{author}{{Cordes}, J.~M.} \& \bibinfo{author}{{Lazio}, T.~J.~W.}
\newblock \bibinfo{title}{{NE2001.I. A New Model for the Galactic Distribution of Free Electrons and its Fluctuations}}.
\newblock \emph{\bibinfo{journal}{arXiv e-prints}} \bibinfo{pages}{astro--ph/0207156} (\bibinfo{year}{2002}).

\bibitem{curran2010}
\bibinfo{author}{{Curran}, P.~A.}, \bibinfo{author}{{Evans}, P.~A.}, \bibinfo{author}{{de Pasquale}, M.}, \bibinfo{author}{{Page}, M.~J.} \& \bibinfo{author}{{van der Horst}, A.~J.}
\newblock \bibinfo{title}{{On the Electron Energy Distribution Index of Swift Gamma-ray Burst Afterglows}}.
\newblock \emph{\bibinfo{journal}{\apjl}} \textbf{\bibinfo{volume}{716}}, \bibinfo{pages}{L135--L139} (\bibinfo{year}{2010}).

\bibitem{gao2013}
\bibinfo{author}{{Gao}, H.}, \bibinfo{author}{{Lei}, W.-H.}, \bibinfo{author}{{Zou}, Y.-C.}, \bibinfo{author}{{Wu}, X.-F.} \& \bibinfo{author}{{Zhang}, B.}
\newblock \bibinfo{title}{{A complete reference of the analytical synchrotron external shock models of gamma-ray bursts}}.
\newblock \emph{\bibinfo{journal}{\nar}} \textbf{\bibinfo{volume}{57}}, \bibinfo{pages}{141--190} (\bibinfo{year}{2013}).

\bibitem{barniolduran2013}
\bibinfo{author}{{Barniol Duran}, R.}, \bibinfo{author}{{Nakar}, E.} \& \bibinfo{author}{{Piran}, T.}
\newblock \bibinfo{title}{{Radius Constraints and Minimal Equipartition Energy of Relativistically Moving Synchrotron Sources}}.
\newblock \emph{\bibinfo{journal}{\apj}} \textbf{\bibinfo{volume}{772}}, \bibinfo{pages}{78} (\bibinfo{year}{2013}).

\bibitem{laskar2019}
\bibinfo{author}{{Laskar}, T.} \emph{et~al.}
\newblock \bibinfo{title}{{A Reverse Shock in GRB 181201A}}.
\newblock \emph{\bibinfo{journal}{\apj}} \textbf{\bibinfo{volume}{884}}, \bibinfo{pages}{121} (\bibinfo{year}{2019}).

\bibitem{ho2019}
\bibinfo{author}{{Ho}, A. Y.~Q.} \emph{et~al.}
\newblock \bibinfo{title}{{AT2018cow: A Luminous Millimeter Transient}}.
\newblock \emph{\bibinfo{journal}{\apj}} \textbf{\bibinfo{volume}{871}}, \bibinfo{pages}{73} (\bibinfo{year}{2019}).

\bibitem{bright2022}
\bibinfo{author}{{Bright}, J.~S.} \emph{et~al.}
\newblock \bibinfo{title}{{Radio and X-Ray Observations of the Luminous Fast Blue Optical Transient AT 2020xnd}}.
\newblock \emph{\bibinfo{journal}{\apj}} \textbf{\bibinfo{volume}{926}}, \bibinfo{pages}{112} (\bibinfo{year}{2022}).

\bibitem{ho2022}
\bibinfo{author}{{Ho}, A. Y.~Q.} \emph{et~al.}
\newblock \bibinfo{title}{{Luminous Millimeter, Radio, and X-Ray Emission from ZTF 20acigmel (AT 2020xnd)}}.
\newblock \emph{\bibinfo{journal}{\apj}} \textbf{\bibinfo{volume}{932}}, \bibinfo{pages}{116} (\bibinfo{year}{2022}).

\bibitem{andreoni2022}
\bibinfo{author}{{Andreoni}, I.} \emph{et~al.}
\newblock \bibinfo{title}{{A very luminous jet from the disruption of a star by a massive black hole}}.
\newblock \emph{\bibinfo{journal}{\nat}} \textbf{\bibinfo{volume}{612}}, \bibinfo{pages}{430--434} (\bibinfo{year}{2022}).

\bibitem{hunter2007}
\bibinfo{author}{Hunter, J.~D.}
\newblock \bibinfo{title}{Matplotlib: A 2d graphics environment}.
\newblock \emph{\bibinfo{journal}{Computing in Science \& Engineering}} \textbf{\bibinfo{volume}{9}}, \bibinfo{pages}{90--95} (\bibinfo{year}{2007}).

\bibitem{harris2020}
\bibinfo{author}{Harris, C.~R.} \emph{et~al.}
\newblock \bibinfo{title}{Array programming with {NumPy}}.
\newblock \emph{\bibinfo{journal}{Nature}} \textbf{\bibinfo{volume}{585}}, \bibinfo{pages}{357--362} (\bibinfo{year}{2020}).
\newblock \urlprefix\url{https://doi.org/10.1038/s41586-020-2649-2}.

\bibitem{virtanen2020}
\bibinfo{author}{Virtanen, P.} \emph{et~al.}
\newblock \bibinfo{title}{{{SciPy} 1.0: Fundamental Algorithms for Scientific Computing in Python}}.
\newblock \emph{\bibinfo{journal}{Nature Methods}} \textbf{\bibinfo{volume}{17}}, \bibinfo{pages}{261--272} (\bibinfo{year}{2020}).

\bibitem{mckinney2010}
\bibinfo{author}{{W}es {M}c{K}inney}.
\newblock \bibinfo{editor}{{S}t\'efan van~der {W}alt} \& \bibinfo{editor}{{J}arrod {M}illman} (eds) \emph{\bibinfo{title}{{D}ata {S}tructures for {S}tatistical {C}omputing in {P}ython}}.
\newblock (eds \bibinfo{editor}{{S}t\'efan van~der {W}alt} \& \bibinfo{editor}{{J}arrod {M}illman}) \emph{\bibinfo{booktitle}{{P}roceedings of the 9th {P}ython in {S}cience {C}onference}}, \bibinfo{pages}{56 -- 61} (\bibinfo{year}{2010}).

\bibitem{pandas2020}
\bibinfo{author}{pandas~development team, T.}
\newblock \bibinfo{title}{pandas-dev/pandas: Pandas} (\bibinfo{year}{2020}).
\newblock \urlprefix\url{https://doi.org/10.5281/zenodo.3509134}.

\bibitem{laskar_meerkat_gcn}
\bibinfo{author}{{Laskar}, T.} \emph{et~al.}
\newblock \bibinfo{title}{{GRB 221009A: MeerKAT detection}}.
\newblock \emph{\bibinfo{journal}{GRB Coordinates Network}} \textbf{\bibinfo{volume}{32740}} (\bibinfo{year}{2022}).

\bibitem{niinuma_jvlbi_gcn}
\bibinfo{author}{{Niinuma}, K.}, \bibinfo{author}{{Yonekura}, Y.}, \bibinfo{author}{{Fujisawa}, K.}, \bibinfo{author}{{Motogi}, K.} \& \bibinfo{author}{{Iwakiri}, W.}
\newblock \bibinfo{title}{{GRB 221009A: Japanese VLBI Network observation.}}
\newblock \emph{\bibinfo{journal}{GRB Coordinates Network}} \textbf{\bibinfo{volume}{32949}} (\bibinfo{year}{2022}).

\bibitem{lacy2020}
\bibinfo{author}{{Lacy}, M.} \emph{et~al.}
\newblock \bibinfo{title}{{The Karl G. Jansky Very Large Array Sky Survey (VLASS). Science Case and Survey Design}}.
\newblock \emph{\bibinfo{journal}{\pasp}} \textbf{\bibinfo{volume}{132}}, \bibinfo{pages}{035001} (\bibinfo{year}{2020}).

\bibitem{condon1998}
\bibinfo{author}{{Condon}, J.~J.} \emph{et~al.}
\newblock \bibinfo{title}{{The NRAO VLA Sky Survey}}.
\newblock \emph{\bibinfo{journal}{\aj}} \textbf{\bibinfo{volume}{115}}, \bibinfo{pages}{1693--1716} (\bibinfo{year}{1998}).

\bibitem{atri_gcn}
\bibinfo{author}{{Atri}, P.} \emph{et~al.}
\newblock \bibinfo{title}{{High-precision position of the compact radio counterpart to GRB221009A.}}
\newblock \emph{\bibinfo{journal}{GRB Coordinates Network}} \textbf{\bibinfo{volume}{32907}} (\bibinfo{year}{2022}).

\end{thebibliography}

\newpage
\bmhead{Supplementary Material}

\setcounter{page}{1}
\renewcommand{\figurename}{Supplementary Figure}
\renewcommand{\tablename}{Supplementary Data Table}
\setcounter{figure}{0}
\setcounter{section}{0}
\setcounter{table}{0}

\begin{figure}[b]
\centering
\includegraphics[width=0.9\textwidth]{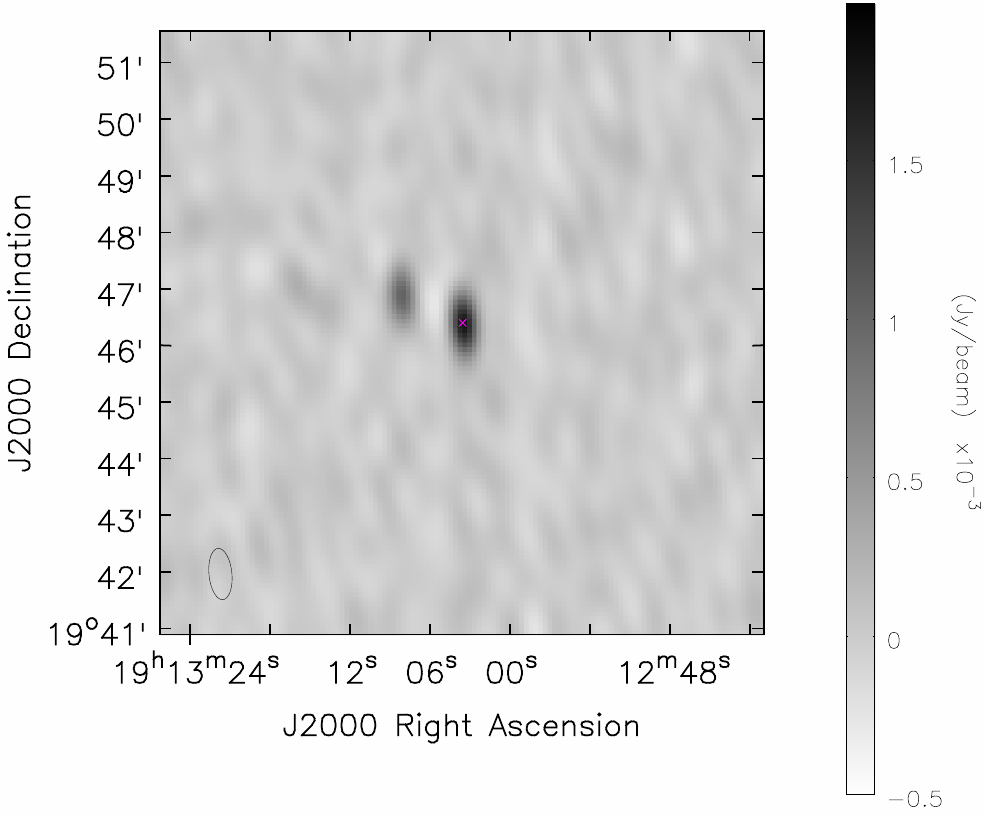}
\caption{\textbf{An example image of GRB 221009A taken with the Arcminute Microkelvin Imager Large Array.} The field of GRB 221009A observed with the AMI--LA. The pink cross marks the position of GRB 221009A (at the pointing centre) from which significant radio emission can be seen. Two field sources are evident to the west of the phase centre, corresponding to those discussed in the Archival Radio Observations section of the Methods and which are shown in Extended Data Figure \ref{fig:archival_radio}. The RMS noise in the image is $\approx60\,\mu\rm{Jy}/\rm{beam}$. The image range is between -0.5\,mJy and 2\,mJy, as indicated by the colour bar. The entire AMI--LA field of view is not shown. The restoring beam for this observation has a major and minor axis of $55''$ and $24''$, respectively, at a position angle of $5.6^{\circ}$.}\label{fig:ami_field}
\end{figure}

\newpage

\begin{figure}[H]
\centering
\includegraphics[width=0.9\textwidth]{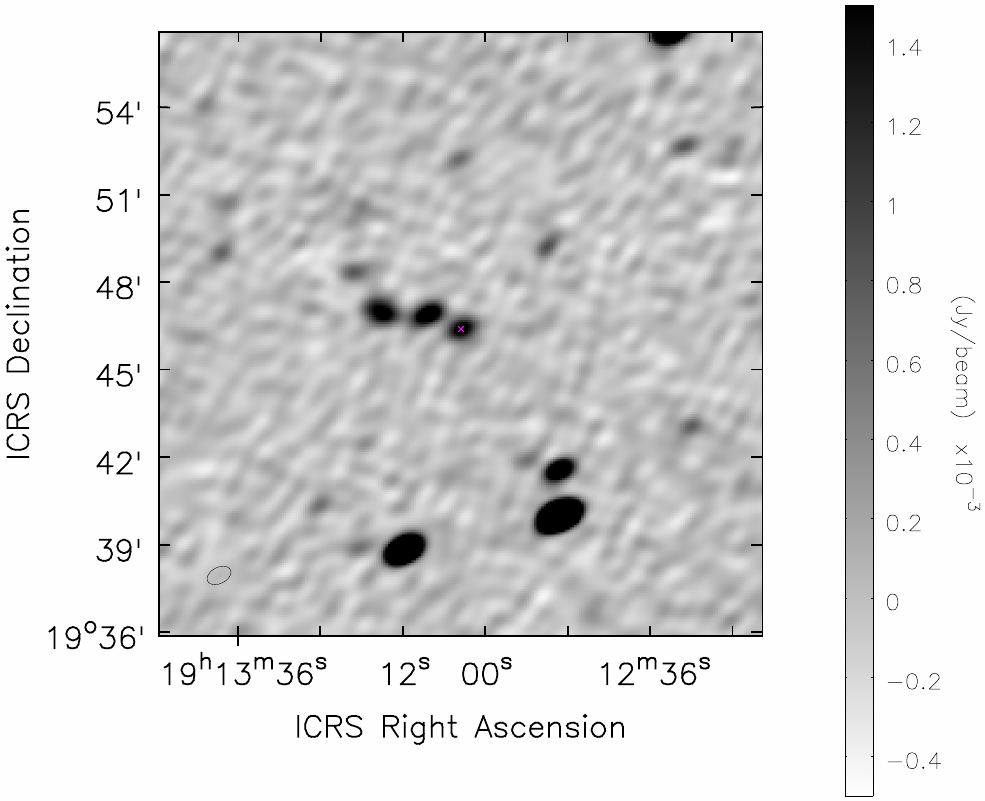}
\caption{\textbf{An example image of GRB 221009A taken with the Allen Telescope Array.} The field of GRB 221009A observed with the ATA at $5\,\rm{GHz}$. The pink cross marks the position of GRB 221009A (at the pointing centre) from which significant radio emission can be seen. Two field sources are evident to the west of the phase centre (in addition to a number of other sources), correspond to those discussed in the Archival Radio Observations section of the Methods and are shown in Extended Data Figure \ref{fig:archival_radio}. The RMS noise in the image is $\approx100\mu\rm{Jy}$ in a source-free region. The image range is between -0.5\,mJy and 1.5\,mJy, as indicated by the colour bar. The whole ATA field of view is not shown. The restoring beam for this observation has a major and minor axis of $52''$ and $32''$, respectively, at a position angle of $116.9^{\circ}$.}\label{fig:ata_field}
\end{figure}

\newpage

\begin{figure}[H]
\centering
\includegraphics[width=0.9\textwidth]{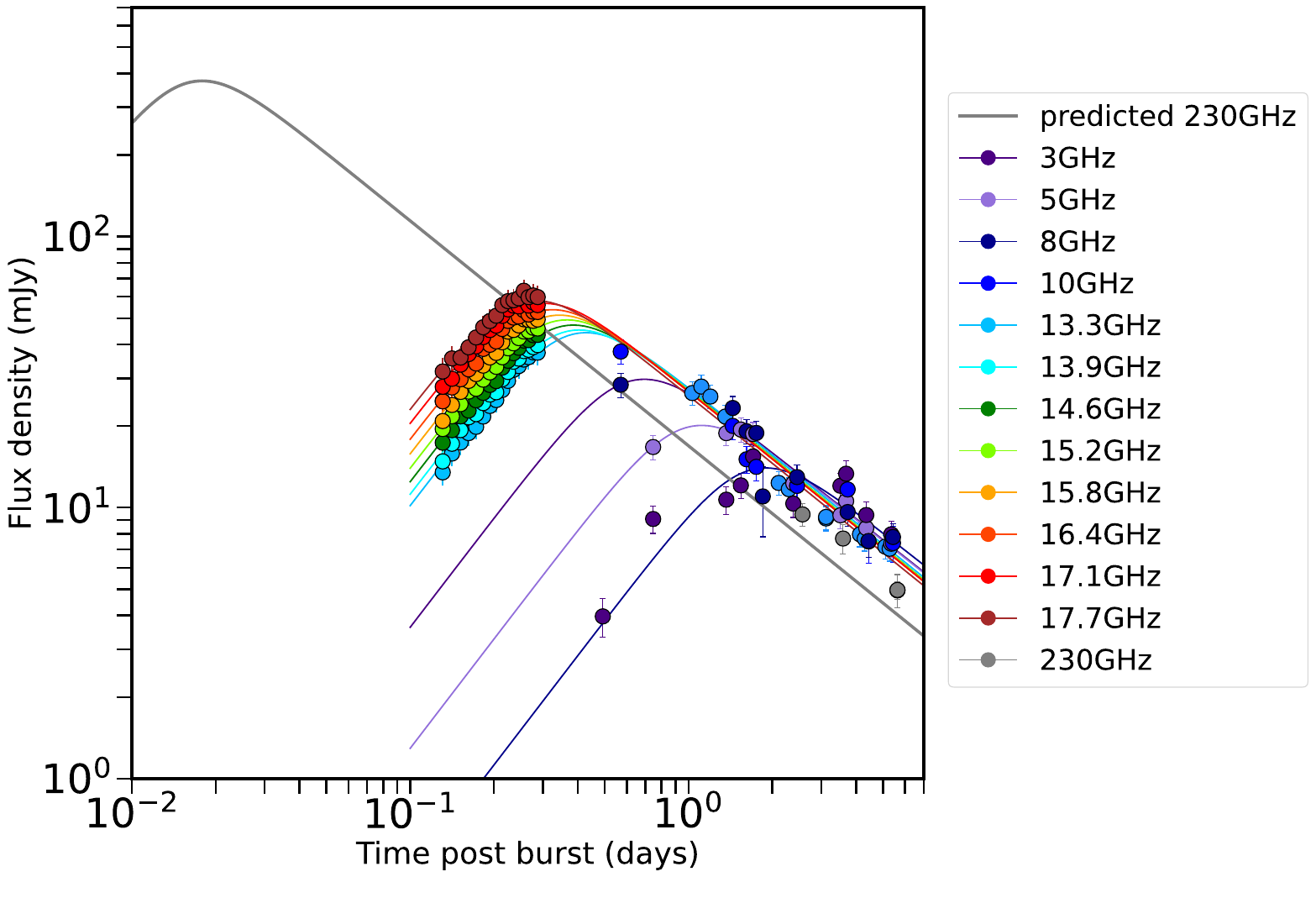}
\caption{\textbf{The predicted peak of sub-mm emission from GRB 221009A.} The predicted $230\,\rm{GHz}$ light curve from the reverse shock of GRB 2210109A based on the peaks observed with the AMI--LA, ATA, and \textit{e}-MERLIN. The $230\,\rm{GHz}$ emission peaks at $\approx0.4\,\rm{hr}$ post burst at a flux density of $\approx370\,\rm{mJy}$. Due to the close proximity of GRB 221009A we do not correct the flux or timescales resulting from the cosmological redshift. The exceptionally bright early-time emission motivates rapid (sub-)mm follow-up even for sources at significant redshifts. Error bars represent $1\sigma$ uncertainties.}
\label{fig:submm_pred}
\end{figure}

\newpage

\begin{figure}[H]
\centering
\includegraphics[width=0.9\textwidth]{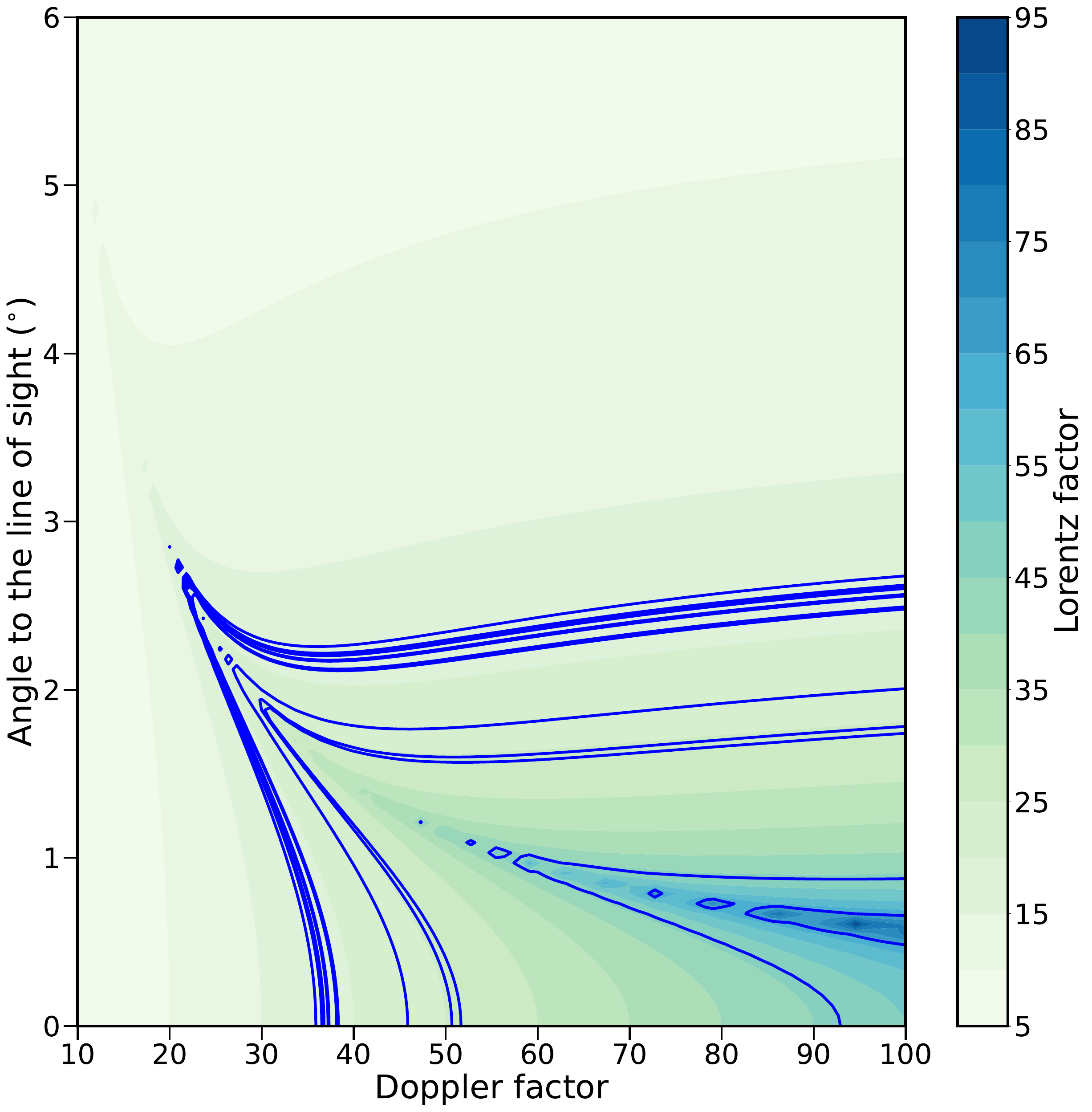}
\caption{\textbf{Constraining the line of sight angle and Doppler factor of the jet from GRB 221009A based on an equipartition analysis of the self-absorbed peak of the reverse shock.} A contour plot of the jet's bulk Lorentz factor as a function of jet angle to the line of sight and relativistic Doppler factor. Overlaid is a series of blue solid lines, each line corresponds to bulk Lorentz factor lower limits obtained from the equipartition analysis (see Methods).}
\label{fig:lf_contour}
\end{figure}

\newpage

\begin{table}[h]
\begin{center}
\begin{minipage}{\textwidth}
\centering
\caption{AMI--LA observations of GRB 221009A between $T_{0}+3.1\,\rm{h}$ and $T_{0}+7.1\,\rm{h}$ measured directly from the complex visibilities per $15\,\rm{min}$ interval and per quick-look frequency channel. A 10\% calibration error should be added in quadrature to the errors reported here. A full machine-readable table can be found as Supplementary Data files as part of the online material.}\label{tab:ami_short}%
\begin{tabular}{@{}cccc@{}}
\toprule
Centroid MJD & Flux Density & Flux Density Error & Frequency \\\newline
 [d] & [mJy] & [mJy] & [GHz] \\
\midrule
59861.6843 & 13.49 & 0.42 & 13.3 \\
59861.6947 & 15.83 & 0.39 & 13.3 \\
... & ... & ... & ... \\
59861.6843 & 14.81 & 0.31 & 13.9 \\
59861.6947 & 17.20 & 0.31 & 13.9 \\
... & ... & ... & ... \\
59861.6843 & 17.36 & 0.35 & 14.6 \\
59861.6947 & 19.30 & 0.34 & 14.6 \\
... & ... & ... & ... \\
59861.6843 & 19.50 & 0.31 & 15.2 \\
59861.6947 & 21.74 & 0.30 & 15.2 \\
... & ... & ... & ... \\
59861.6843 & 20.86 & 0.34 & 15.8 \\
59861.6947 & 23.92 & 0.32 & 15.8 \\
... & ... & ... & ... \\
59861.6843 & 24.66 & 0.37 & 16.4 \\
59861.6947 & 27.74 & 0.36 & 16.4 \\
... & ... & ... & ... \\
59861.6843 & 27.91 & 0.47 & 17.1 \\
59861.6947 & 29.88 & 0.44 & 17.1 \\
... & ... & ... & ... \\
59861.6843 & 31.79 & 2.14 & 17.7 \\
59861.6947 & 35.52 & 1.39 & 17.7 \\
... & ... & ... & ... \\
\botrule
\end{tabular}
\end{minipage}
\end{center}
\end{table}

\newpage

\begin{table}[h]
\begin{center}
\begin{minipage}{\textwidth}
\centering
\caption{Flux density ($F_{\nu}(t)\propto t^{\alpha}$) and spectral ($\nu(t)\propto t^{\alpha}$) evolution for regions of parameter space relevant to our early time radio observations \cite{vanderhorst2014}. The two columns describe a thin and thick shell, respectively, for $p>1$ and in the post shell crossing regime. Note these scenarios do not consider the pre-shock crossing phase.}\label{tab:relevant_scalings}%
\begin{tabular}{@{}ccc@{}}
\toprule
 & $\alpha_{\rm{thin}, p>1, post-cross}$ & $\alpha_{\rm{thick}, p>1, post-cross}$ \\
\midrule
$F_{(\nu_{m} < \nu < \nu_{sa})}$ & $\frac{5(5g+8)}{14(2g+1)}$ & $\frac{113-22k}{24(4-k)}$\\
$F_{(\nu_{m} < \nu_{sa} < \nu)}$ & $-\frac{3p(5g+8)+7g}{14(2g+1)}$ & $-\frac{p(73-14k)+3(7-2k)}{24(4-k)}$\\
$F_{\nu_{sa}}$ & $-\frac{5g(5p+6)+20(2p+1)}{7(2g+1)(p+4)}$ & $\frac{-2k(12p+13)+126p+109}{12(k-4)(p+4))}$ \\
$\nu_{sa}$ & $-\frac{3p(5g+8)+8(4g+5)}{7(2g+1)(p+4)}$ & $-\frac{p(73-14k) + 2(67-14k)}{12(4-k)(4+p)}$ \\
\botrule
\end{tabular}
\end{minipage}
\end{center}
\end{table}

\newpage


\end{document}